# Phase Field Theory of Nucleation and Polycrystalline Pattern Formation

**László Gránásy, Tamás Pusztai, Tamás Börzsönyi**

*Research Institute for Solid State Physics and Optics, Budapest, Hungary*

## CONTENTS

## 1. INTRODUCTION

Crystalline materials play an essential role in our everyday life. Most of them are polycrystalline, i.e., composed of a large number of crystallites, whose size, shape and composition distributions determine their properties and failure characteristics [1]. The size scale of the constituent crystal grains varies between a few nanometers (nanocrystalline alloys) and centimeters in different classes of materials. Despite intensive research, many aspects of the formation of polycrystalline matter are poorly understood. The main source of difficulties is the process of nucleation – the least understood early stage of crystallization – during which crystallites capable for further growth form via thermal fluctuations. While nucleation takes place on the nanometer scale, its influence extends to other size scales. Controlled nucleation [2] might be a tool for tailoring the microstructure of matter for specific applications. The complexity of polycrystalline freezing is especially obvious in the case of thin (few times 10 nm) polymer layers, which show an enormous richness of crystallization morphologies. These quasi two-dimensional structures give important clues to the mechanisms that govern the formation of polycrystalline patterns. While polycrystalline pattern formation plays an important role in classical materials science and nanotechnology, it has biological relevance as well (e.g., the appearance of semi-crystalline spherulites of amyloid fibrils is associated with the Alzheimer and Creutzfeldt-Jakob diseases, type II. diabetes, and a range of systemic and neurotic disorders [3]).

The polycrystalline morphologies observed in nature, laboratory and technology might be somewhat arbitrarily divided into two categories:

(a) *Foam-like structures* that form by the impingement of independently nucleated single crystals. These structures are characteristic to equiaxed growth in cast materials.

(b) *Polycrystalline growth forms*, which form by the nucleation of new grains at the solidification front.

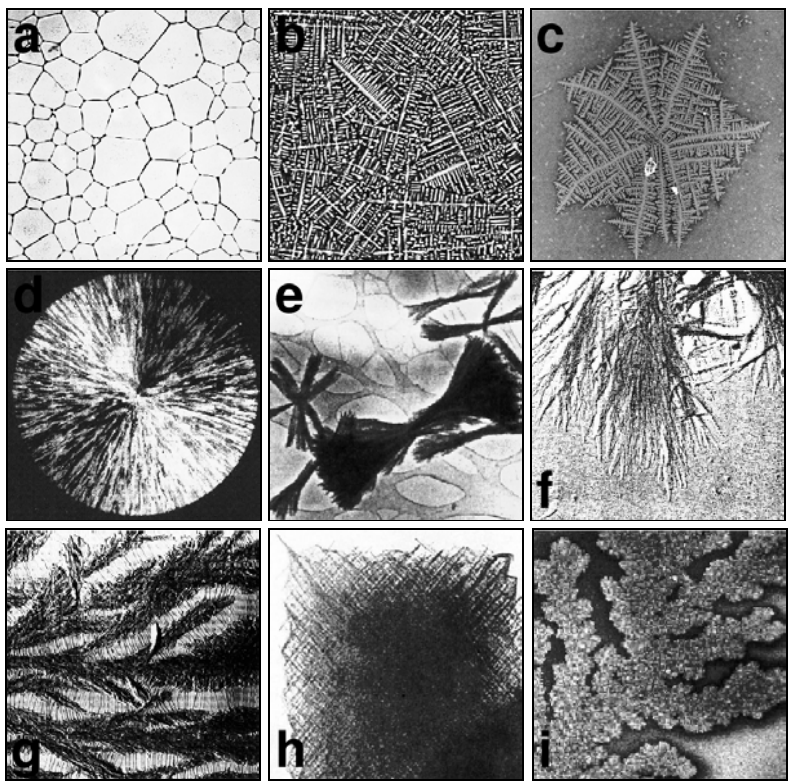

**Figure 1.** Polycrystalline microstructures. (a) Foam-like morphology formed by competing nucleation and growth [4]. (b) Polycrystalline dendritic structure formed by competing nucleation and growth in the oxide glass $(ZnO)_{61.4} \cdot (B_2O_3)_{38.6} \cdot (ZnO_2)_{28}$ [5]. (c) 'Dizzy' dendrite formed in clay filled polymethyl methacrylate-polyethylene oxide thin film (for experimental conditions see [6]). (d) Spherulite formed in pure Se (reproduced from [7] with permission of Elsevier). (e) Crystal sheaves in pyromellitic dianhydrite-oxydianilin poly(imid) layer (reproduced from [8] with permission of the American Chemical Society). (f) Arboresque growth form in polyglycine (reproduced from [9] with permission of the American Institute of Physics). (g) Polyethylene spherulite crystallized in the presence of n-paraffin [10]. (h) 'Quadrite' formed by nearly rectangular branching in isotactic polypropylene [11]. (i) Fractal-like polycrystalline aggregate of electrodeposited Cu (reproduced from [12] with the permission of Nature Publishing Group).

Typical polycrystalline microstructures are displayed in Fig. 1. The foam-like structure formed by the impingement of individual single crystals is shown in Fig. 1(a) [4]. A polycrystalline dendritic morphology formed during chemical-diffusion-controlled anisotropic crystal growth in an oxide glass is presented in Fig. 1(b) [5]. *Polycrystalline growth forms* are shown in Figs. 1(c)-(i) [6 – 12]. Recent experiment on polymer films revealed that particulate additives may transform the single crystal dendrites into disordered polycrystalline dendrites [Fig. 1(c)] [6]. *Spherulites* [Fig. 1(d)] provide a classic example of polycrystalline growth. This structure has been observed in a wide range of materials including pure metals [7], alloys, polymers, minerals, and biological systems. In systems that form fibers during crystallization, the formation of spherulites commonly starts with the appearance of crystal *sheaves* of diverging ends [Fig. 1(e)] [8], which occasionally develop into less space-filling structures [Figs. 1(f) and (g)] [9, 10]. Random crystallographic branching of nearly 90 degree, yields 'quadrites' [Fig. 1(h)] observed in certain polymeric systems [11]. Disorderly growth of small crystallites during electrodeposition often result in irregular, *fractal-like* structures [Fig. 1(i)] [12]. While the specific mechanisms that lead to the formation of these complicated structures are usually poorly understood, it is expected that nucleation, diffusional instabilities, crystal symmetries, and impurities play important roles.

Like the exploration of many other complex problems, research and understanding of polycrystalline matter profit excessively from the improving performance of modern computers. In the past decade, various models were developed and applied to address complex solidification problems. These include the lattice-gas/lattice-Boltzmann/cellular automata models [13 – 20], the level set [21, 22], and other front tracking techniques [23 - 25], and the phase field theory [26 – 33].

Among these approaches, which all have their specific strengths and weaknesses, the phase field theory became perhaps the most popular. This originates partly from the fact that it starts from thermodynamic/statistical mechanical principles and obtains governing equations for microstructure evolution at the end; connecting thus thermodynamic and kinetic properties with microstructure via a mathematical formalism. Since many of the developments have been reviewed recently [26 – 33], here we briefly outline the phase field concept, and review only the latest developments that demonstrate the applicability of this approach for polycrystalline solidification, and refer to previous work only to the level required to place these developments into context. Our review covers recent advances made in modeling nucleation and polycrystalline solidification in various systems including morphology evolution in quasi-two-dimensional layers, the formation of spherulites, transformation kinetics, interaction of particulate additives with crystallization, and the evolution of nucleation patterns on surfaces.

We start with a brief introduction of the phase field method (Sec. 2.1). Early models of multi-particle solidification are reviewed in the subsequent paragraph (Sec. 2.2). Recent approaches, which rely on the use of an orientation field to distinguish crystallites with different crystallographic orientations are described in Sec. 2.3. The theory is then used to calculate the height of the nucleation barrier and the nucleation rate (Sec. 2.4). We present a quantitative test of theory, performed for the well-known hard-sphere system, where all the parameters the phase field theory needs can be fixed with a high accuracy via results from atomistic simulations. Next (Sec. 2.5), we deal with polycrystalline morphologies that appear in ideal and regular solutions, and review the kinetics of polycrystalline freezing in such systems. The formation of polycrystalline growth morphologies (spherulites, disordered dendrites, fractal-like aggregates, etc.) characteristic to far-from-equilibrium freezing is explored in Sec. 2.6. Here, we identify the essential factors that govern polycrystalline solidification. Patterns from heterogeneous nucleation on walls and foreign particles, and crystallization in confined space (in porous matter or in channels) are addressed in Sec. 2.7. Finally (Sec. 3), we call attention to a few promising approaches that may set the future trends in this branch of computational materials science.

# 2. PHASE FIELD THEORY OF NUCLEATION

In this section, we review models developed for describing single crystal patterns, attempts to extend them to polycrystalline freezing, and their applications to complex polycrystalline morphologies observed in the laboratory and nature.

## 2.1 Phase Field Theory of Crystal Growth

### 2.1.1. The Phase Field Method

The phase field theory [26 – 33] is a descendant of the van der Waals/Cahn-Hilliard/Landau type classical field theoretic approaches [34 – 38]. It originates from a single-order parameter gradient theory of Langer from 1978 [39]. Similar models were independently developed by Collins and Levine [40] and Caginalp [41]. In the phase field theory, the local state of matter is characterized by a non-conserved structural order parameter $\phi(\mathbf{r},t)$, called phase field, which monitors the transition between the solid and liquid states. It can be viewed as a structural order parameter that measures local crystallinity. It is also interpreted as the volume fraction of the crystalline phase. In the presence of $n$ crystalline phases, and one disordered phase a minimum of $n$ phase fields are needed $\{\phi_i(\mathbf{r},t)\}$. In some models, such as the multi-phase field theory by Steinbach *et al.* [42], a separate phase field is introduced for every crystal grain. This may lead to thousands of phase fields when addressing multi-grain problems. While these multi-phase field theories are very powerful methods for describing complex morphologies, the inclusion of thermal fluctuations and hence a physical modeling of nucleation is not straightforward.

In the course of developing the model, one "expands" the free energy density (or entropy density) of the inhomogeneous system (liquid + solid phase(s)) in terms of the structural order parameter(s) $\{\phi_i\}$, the chemical composition field(s) $\{c_i\}$, the orientation field, etc., and their spatial derivatives, retaining only those spatial derivatives that are allowed by symmetry considerations. The free energy of the system is a local functional of the field variables, and can usually be regarded as a specific case of the general functional below:

$$F = \int d\mathbf{r} \left\{ \sum_{i,j} \alpha_{ij} \left( \nabla \phi_i \nabla \phi_j \right) + \sum_{i,j} \beta_{ij} \left( \nabla c_i \nabla c_j \right) + \sum_{i,j} \gamma_{ij} \left( \nabla \phi_i \nabla c_j \right) ... + f\left[ \{\phi_i\}, \{c_i\}, T, ...\right] \right\} \ . \tag{1}$$

(In more complex formulations higher order differential operators are also used.) The gradient terms that penalize the spatial change of the fields give rise to the interfacial energies, and lead to diffuse interfaces as opposed with the mathematically sharp interfaces of the classical models. The coefficients $\alpha_{ij}$, $\beta_{ij}$ and $\gamma_{ij}$ may depend on temperature, orientation, and the field variables. The bulk free energy density $f(\{\phi_i\}, \{c_i\}, T, ...)$ has two or more minima corresponding to the bulk liquid and crystalline phases. It is worth mentioning that the free energy functional of solid-liquid systems can be derived on physical grounds using the density functional approach [43, 44], and after appropriate simplifications it can be cast into the form of gradient theory. These molecular theories are, however, usually too complicated to address complex solidification morphologies. Accordingly, most approaches rely on phenomenological free energy (or entropy) functionals whose form owes much to the Landau model of phase transitions [37, 38]. The phase field approaches usually differ in the field variables considered, and the actual form of coupling between the fields. Once the free energy functional is defined, the formalism that describes dynamics under non-equilibrium conditions follows almost automatically, though further approximations are usually made.

Starting from the principle of positive entropy production (or decreasing free energy), partial differential equations are derived for the evolution of the phase field and the other field variables [26 − 33, 39 − 53]. The governing equations differ for non-conserved fields (whose spatial integral varies with time during the transition, e.g., phase field) and conserved fields (whose spatial integral is constant, e.g., chemical composition):

$$\text{Non-conserved fields:} \quad \dot{\phi}_i = -M_{\phi_i} \frac{\delta F}{\delta \phi_i} + \zeta_i$$

$$\text{Conserved fields:} \quad \dot{c}_i = \nabla \left\{ M_{c_i} \nabla \frac{\delta F}{\delta c_i} \right\} - \nabla \zeta_{j_c}$$

The appropriate field mobilities $M_i$ set the typical time scale for the evolution of respective fields. Here the simplifying assumption has been made that there are no mobility cross couplings between the applied fields. To mimic the thermal fluctuations, Gaussian white noise terms $\zeta_i$ (random current applies for conserved quantities) of amplitudes determined by the fluctuation-dissipation theorem are added to the governing equations [38, 49, 54 − 56]. The time evolution of the non-conserved fields is coupled to those of the conserved fields (i.e., the phase field model can be regarded as a generalized Hohenberg-Halperin Model C-type classical field theory [37]). These governing equations are highly non-linear, and capture such phenomena as diffusional instabilities, enabling the approach to describe complex solidification patterns including thermal dendrites [46 − 48, 53, 57, 58] and solutal (chemical diffusion controlled) dendrites [50, 51, 59 − 62] (Fig. 2), eutectic [49, 63 − 67], monotectic [68] and peritectic solidification [69 − 71], banded structures [72], and many others. (Remarkably, various formulations of the phase field theory for thermal dendrites lead to comparable dynamics [73].)

While the phase field models provided *qualitative understanding* to such phenomena, as the evolution of morphological instabilities, solute trapping, etc., *quantitative* prediction of microstructure, which has a vast practical importance in optimizing and designing materials for specific applications, remains a major challenge. The main difficulty quantitative phase field modeling has to face is that sub-nanometer spatial resolution is needed in the interfacial region, which extends to a couple of nanometers according to experiment [74, 75] and computer simulations [76 − 77]. Indeed, the phase field approaches recover the diffuse interfaces by introducing square-gradient terms in the free energy penalizing sharp changes of the fields. However, the interface thickness is usually orders of magnitude smaller than the objects of interest, thus numerical solution of the equations, at the resolution required to describe the nanometer thick diffuse interfaces properly, is, as yet, impossible (in two and higher dimensions) even with the most powerful computers. Accordingly, accurate simulations are limited to small volumes and short times. In the case of larger systems, we might satisfy ourselves with qualitative modeling that relies on unphysically broad interface, i.e., the interface thickness is regarded as model parameter. This, however, may influence the growth rate, the composition of the solid (solute trapping), and other features. To overcome this difficulty, methods have been worked out to ensure the proper interface dynamics by adjusting the model parameters and introducing interface currents (i.e., new term in the phase field equations) to compensate for the unphysical effects of a too thick interface [57, 79 – 81]. These methods make a *quantitative* phase field modeling of dendritic solidification feasible for thermal dendrites and dendrites in dilute solutions [57, 58, 82]. While quantitative modeling of such dendrites with model parameters deduced from atomistic simulations is one of the most spectacular successes of the theory, a generally applicable approach has yet to be developed.

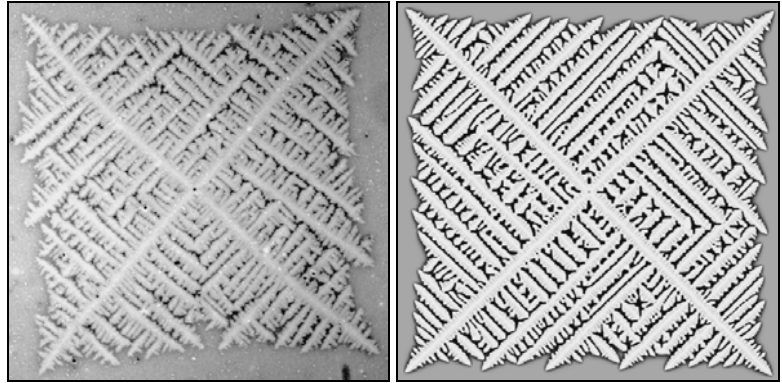

**Figure 2.** Single crystal dendrites in polymethyl methacrylate-polyethylene oxide film (reproduced from [195], © 2003, with the permission of the Nature Publishing Group) and in a phase field simulation performed on a 2000 × 2000 rectangular grid (26.3 μm × 26.3 μm) at 1574 K and supersaturation $S = (c_1 - c_\infty)/(c_1 - c_s) = 0.80$ using the thermodynamic and interfacial properties of Ni-Cu, and a 15% anisotropy for the interfacial free energy. Here $c_\infty$, $c_1$, and $c_s$ are the initial composition of the liquid, and the liquidus and solidus compositions. (For details see [195].) The simulation has been made using Model A (see Appendix).

Systems, in which the interfacial free energy and/or kinetic coefficient have strong anisotropies, and faceted morphologies expected to occur represent another important challenge to quantitative phase field modeling [83 – 88]. A further difficulty associated with quantitative phase field calculations is that the detailed information on the system needed for such computations, such as the magnitude and anisotropy of the phase field mobility and the interfacial free energy, are generally inaccessible. Linking the phase field theory with atomistic simulations, and the evaluation of the parameters of the phase field theory (mobility, anisotropies, interfacial free energy) is a possible resolution of this problem [32, 89 – 92].

We emphasize that although the phase field theory is a phenomenological model, similar models can be derived on physical grounds using the density functional theory. Considering the crystal as a highly inhomogeneous liquid, whose number density $\rho(\mathbf{r})$ peaks at the lattice sites, the Fourier amplitudes of the number density distribution $\rho(\mathbf{r})$ can be regarded as structural order parameters representing crystalline atomic order [43, 44]. The number of these structural order parameters can be reduced if the density peaks at the atomic sites are assumed to have a Gaussian form: Under such conditions, all Fourier amplitudes of the number density can be expressed uniquely in terms of the amplitude of a dominant density wave, therefore, a single structural order parameter is sufficient for the description [93, 94]. Thus, the phase field can be viewed as the amplitude of the dominant Fourier component of the $\rho(\mathbf{r})$ representing the crystal. Shih *et al.* [95] developed another method to obtain a free energy functional consistent with crystal symmetries on the basis of a Landau expansion, an approach that has been applied to crystal nucleation and growth by Iwamatsu and Horii [96, 97]. This route offers physical interpretation for the model parameters, and derivation of the functions introduced intuitively into the phase field theory. Formulation of a single order parameter theory for crystal nucleation in systems with bcc and fcc structure has recently been developed along this line [98].

Note that the models reviewed above – with the exception of the multi phase field approach – are unable to address anisotropic growth of crystal grains with different crystallographic orientation.

Due to the practical importance of polycrystalline materials, extensive efforts have been made to extend the phase field approach to this case.

### 2.1.2. Technical Issues

The solution of the phase field equations usually represents an extremely demanding computational task. Many of the real metallurgical problems are inaccessible for quantitative phase field simulations simply due to the shear extent of the numerical work required. Therefore, efficiency of the numerical algorithms applied is an important issue. While the improving performance of computers will certainly ease the situation in time, there are several methods that can offer immediate help to overcome these limitations:

***Parallel computing:*** The simulation is cut into parts, which individual parts are run on different computers, that communicate the relevant results to each other time to time. Such approach has been applied recently to a variety of phase field problems [60,99 – 103] allowing simulations on grids as large as $10^4 \times 10^4$ and the handling of hundreds of dendritic particles. Details of parallel computing, applied to phase field simulations in 2D and 3D, are presented by George and Warren [60].

***Adaptive mesh:*** The number of mesh points used in the simulations can be drastically reduced by the adaptive mesh techniques that use high resolution only in the vicinity of interfaces. Description of such procedures can be found in several recent works [102, 104 – 110] .

***Random-walk algorithms:*** Problems, where long-range diffusion fields play a role, can be efficiently handled using multiscale random-walk algorithms [111, 112].

***Spectral methods:*** Fourier methods combined with operator splitting, might lead to algorithms of far improved performance. For example, in the case of the Cahn-Hilliard type problems, spectral methods are known to increase the computational speed/accuracy enormously [113, 114]. Popov *et al.* [115] applied Fourier methods to 2D phase field problems recently.

***Lattice anisotropy:*** Due to the rectangular lattice usually employed, the patterns that form are usually anisotropic even if all physical properties are isotropic. Lattice anisotropy in the simulations is suppressed by different methods. Smaller spatial steps reduce the lattice anisotropy, thus adaptive grid methods are advantageous. The lattice anisotropy can be reduced by the introduction of kinetic and surface anisotropies that balance the anisotropy from the lattice [57], via the use of hexagonal, random or locally rotated lattices [116, 117], and by isotropic numerical schemes for the differential operators [118, 119].

While some of these methods have already been applied for polycrystalline solidification problems, others might be employed in the nearest future.

## 2.2. Early Models of Polycrystalline Freezing

Polycrystalline morphologies of various complexity are observed, which range between the relatively simple foam-like structures – formed by nucleation, growth, and impingement of roughly spherical crystallites – and the rather complicated, semi-crystalline growth patterns shown in Fig. 1. Their modeling requires approaches of comparably different complexity, as we review below.

### 2.2.1. Formal Theory of Polycrystalline Solidification

*Foam-like grain boundary structures* appear when individually nucleating crystallites impinge upon each other as they grow. The time evolution of the crystalline fraction in such "nucleation – growth" problems is of primary interest for diverse branches of science including materials science, chemistry, atmospheric sciences, geophysics and astronomy, and is usually described in the framework of the Kolmogorov-Johnson-Mehl-Avrami (KJMA) theory (for review see [120]). The central notion of this approach is the 'overlapping' transformed fraction, defined as

$$Y(t) = \frac{4\pi}{3} \int_0^t J(\tau) \left\{ \int_\tau^t v(\vartheta) d\vartheta \right\}^3 d\tau , \qquad (2)$$

where $J$ and $v$ are the nucleation and growth rates, while the integration variables $\vartheta$ and $\tau$ have dimensions of time. During the early stages of the process when the crystallites are far from each other and grow independently, $Y(t)$ coincides with the true crystalline fraction $X$. However, soon these crystalline 'particles' (as predicted by (2)) overlap, multiply covered regions appear, and (2) overestimates the true crystalline fraction. Here the volume of the critical fluctuations is neglected, as apart from the very early stages of the transition, their contribution to the transformed fraction is small relative to the contribution by the supercritical particles. The true crystalline fraction $X$ can be related to $Y$ via a simple mean field expression $dX = (1 − X)\,dY$, which counts only that fraction of $dY$ that falls on the untransformed region. Upon integration, one obtains $X = 1 − exp\{−Y\}$, an expression that is exact if (i) the transformation takes place in an infinite medium; (ii) the new particles nucleate the untransformed region at a rate, which is independent of the coordinates; and (iii) the freely growing particles have the same convex shape and orientation, while the growth rate depends on time through a multiplicative factor common for all directions. For example, assuming constant nucleation and growth rates and an infinite system, the time evolution of the crystalline fraction follows the KJMA scaling $X = 1 − exp\{−(t/t_0)^p\}$, where $t_0$ is a time constant related to the nucleation and growth rates, $p = 1 + d$ is the Avrami-Kolmogorov exponent, and $d$ is the number of dimensions. If the number of nuclei is fixed (which occurs after early-stage site-saturation, or when quenched-in or athermal nuclei dominate), one obtains $p = d$. (Derivation of the KJMA relationship using the time cone method by Cahn is given in [121, 122].) A variety of processes are known to deviate from the KJMA scaling (for example, the nucleation and growth of anisotropic particles [123-126]). A practically important question is whether the KJMA scaling might work in the presence of chemical diffusion. Condition (iii) is evidently violated here, as diffusion-controlled growth yields a growth rate that diminishes with increasing particle size. Although no exact treatment of the problem is available, it has been suggested [120] that under such conditions, $p ≈ 1 + d\,/2$ applies for constant nucleation rate and $p ≈ d/2$ for fixed number of particles. Recent experimental studies [127] in agreement with approximate descriptions [127,128] find, however, deviation from this behavior for diffusion mediated 'soft impingement' of crystal particles. Being able to handle the interacting diffusion fields of growing particles, the phase field theory is an ideal tool to address such problems.

### 2.2.2. Phase Field Models with Isotropic Growth

When solidification takes place with nucleation and isotropic growth of the particles, one does not need to introduce a local crystallographic orientation to address the kinetics of freezing. Jou and Lusk [129] applied a scalar order parameter theory to study the formation of foam-like multigrain structures in one-component, isotropic systems. They observed deviations from a constant growth rate only at short times, and the transformed fraction was found to follow closely the KJMA scaling, except in the case of very large nucleation rates. Elder *et al.* [49] developed a two-field theory (phase and concentration fields) with Langevin-noise to induce crystal nucleation for describing multigrain solidification in isotropic eutectic system. The $p = 3$ they found for the Avrami-Kolmogorov exponent is consistent with the absence of long-range diffusion. (Short-range diffusion, parallel with the growth front, is the dominant diffusion mode here.) Gránásy *et al.* [130] investigated diffusion-controlled solidification in a binary system of ideal solution thermodynamics (Ni-Cu), initiated by randomly positioned supercritical particles of fixed number. They found that the Avrami-Kolmogorov exponent diminishes as the crystallization advances, a behavior that follows the trend seen in experiment [127] and predicted by approximate theory [127,128].

### 2.2.3. Phase Field Models Addressing Anisotropic Growth

To describe the impingement of a large number of crystallites that grow anisotropically [shown in Fig. 1(b)], one needs to incorporate the crystallographic orientation into the theory that determines the preferred growth directions.

The first phase field model that allows for different crystallographic orientations in a solidifying system (Morin *et al.* [131]), relies on a free energy density that has $n$ wells, corresponding to $n$ crystallographic orientations, breaking thus the rotational symmetry of the free energy. In this work, homogeneous nucleation has been mimicked by randomly introducing seeds (in space and time) that exceed the critical size. Simulations have been performed to study polymorphous crystallization, during which the composition of the liquid remains close to that of the crystal, therefore, chemical diffusion plays a minor role, and the KJMA form fits the simulations reasonably well. A weakness of the model is that the rotational invariance of the free energy density had to be sacrificed to introduce a finite number of crystallographic orientations, which form grains with a diffuse interface (grain boundary) between them.

A significantly different approach for addressing the formation of particles with random crystallographic orientations is represented by the multi-phase field models [42, 132 – 135], which offer flexibility at the expense of enhanced complexity. These models have been used to study polycrystalline dendritic and eutectic/peritectic solidification during directional and equiaxed conditions [66, 133, 134]. They have also been successfully applied for describing the time evolution of multigrain structures. However, the large number of phase fields applied in these approaches leads to difficulties, when nucleation is to be initiated by Langevin noise. While noise-induced nucleation can certainly be substituted by inserting the nuclei by 'hand' into the simulations, this procedure becomes excessively non-trivial when structures that require the nucleation of different crystallographic orientations at the growth front [Figs. 1(c)–(i)] are to be addressed.

## 2.3. Phase Field Models with Orientation Field

Since the first model of polycrystalline solidification that incorporates crystallographic orientations and a rotationally invariant free energy (Kobayashi *et al.* [136, 137]) is the basis of later developments, we discuss it here in detail.

In two dimensions, crystallographic orientation can be specified by a single angular variable that sets the tilt of the crystal planes in the laboratory frame. Accordingly, Kobayashi *et al.* [136, 137] introduced a non-conserved scalar orientation field $\theta(\mathbf{r},t)$ that sets the local crystal orientation in the crystallized regions relative to which the angular dependencies of the interfacial free energy and the kinetic coefficient are measured. A heuristic approach is then used to derive the orientational free energy $F_{ori}$. Following the general philosophy of the phase field method, it is assumed that the orientational free energy is a local functional (i.e., it may depend on the field variables and their derivatives). Another requirement is the invariance of the free energy with respect to rotations of the laboratory frame (i.e., explicit dependence on $\theta$ and its powers is excluded). Seeking the orientational free energy in the form of $F_{ori} = \int dr\, H\, |\nabla \theta|^n$, where the constant $H$ and the exponent $n$ have yet to be specified, the free energy of a planar interface between two semi-infinite crystal grains of misorientation $\Delta \theta$ can be expressed as

$$F_{ori} = \int_0^L dz\, H\, |\nabla \theta|^n \;\propto\; \frac{(\Delta \theta)^n}{L^{n-1}} \;. \tag{3}$$

Here $L$ is the thickness of the interface region and the integration is taken with respect to the spatial coordinate $z$ perpendicular to the interface. For $n > 1$, the orientational free energy $F_{ori}$ decreases with increasing interface thickness. Thus the system lowers its free energy by broadening the interface region indefinitely. [In fact, $n = 2$ is also a reasonable model for a grain boundary, with the caveat that real grain boundaries are properly described as a wall of dislocations. Note that dislocations can be regarded as singularities in the $\nabla \theta$ field. Since we are interested here in modeling of coherent lines of dislocations (i.e., grain boundaries), we disregard single dislocations.] Apparently, the most plausible choice that leads to a stable interface with non-zero free energy is $n = 1$. In this case, the free energy contribution associated with the interface is proportional to $\Delta \theta$ [see Fig. 3(a)], provided that $\theta(z)$ is monotonic [if $\theta(z)$ is non-monotonic, the energy is not a minimum]. This minimization, however, leaves the interface profile $\theta(z)$ undefined. This arbitrariness can be remedied assuming that the coefficient $H$ varies with $z$ so that it has a minimum at the interface [see Fig. 3(b)]. Minimization of free energy will then lead to a stepwise variation of $\theta(z)$, a behavior that approximates reasonably the experimental reality of stable, planar grain boundaries. Such a minimum has already been realized making the coefficient $H$ dependent on the phase field, or on an extra "solid order parameter" that determines whether the solid material is crystalline or disordered [136, 137]. Due to the non-analytic nature of the orientational free energy, the equation of motion of the orientation field defines a *singular diffusivity* problem that requires special care when handled numerically [138, 139]. This approach has been applied for describing the impingement of fixed number of anisotropically growing particles of diverse morphologies, including dendritic solidification in single component [136, 137] and binary liquids [102, 140]. Diverse grain boundary related problems, such as grain boundary wetting and grain coarsening in polycrystalline matter via grain boundary migration and rotation, have also been addressed [141 - 144].

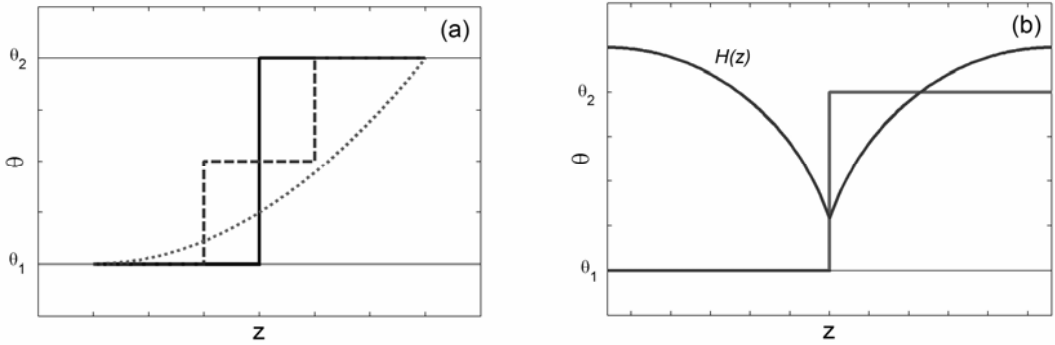

**Figure 3.** (a) $\int dz \, |\nabla\theta| = |\Delta\theta|$ is the same for the three $\theta(z)$ functions (dotted, dashed, and solid lines), since they vary monotonically between the same end points. (b) If the coefficient of $|\nabla\theta|$ has a minimum in the interface — after free energy minimization — the orientation field changes stepwise between the two orientations.

Modeling of nucleation of crystallites with different crystallographic orientations (either in the liquid or at the growth front) requires a further important step made by Gránásy *et al.* [99, 100], who extended the orientation field $\theta$ into the liquid phase, where $\theta$ has been made to fluctuate in time and space. Assigning local crystallographic orientation to liquid regions, even a fluctuating one, may seem artificial at first sight. However, due to geometrical and/or chemical constraints, a short-range order exists even in simple liquids, which is often similar to the one in the solid. Rotating the crystalline first-neighbor shell centered to a liquid molecule so that it aligns optimally with the local liquid structure, one may assign a local orientation to every molecule in the liquid. The orientation obtained in this manner fluctuates in time and space. The correlation of the atomic positions/angles shows how good this fit is. (In the model by Gránásy *et al.* [99, 100], the fluctuating orientation field and the phase field play these roles.) Approaching the solid from the liquid, the orientation gradually becomes more definite (the amplitude of the orientational fluctuations decreases) and eventually matches to that of the solid, while the correlation between the local liquid structure and the crystal structure improves. In this model, called Model A henceforth (for details see the Appendix), the orientation field and the phase field are strongly coupled to recover this behavior. *We emphasize that the models termed Models A to C in this paper differ from Models A to C of the usual Hohenberg-Halperin [37] classification.*

In Model A, the orientational free energy density has the form $f_{ori} = HT\,[1 - p(\phi)]\,|\nabla\theta|$, where $p(\phi)$ is the interpolation function (see Appendix) that varies between 0 and 1, while $\phi$ changes from $\phi = 0$ in the bulk solid phase to $\phi = 1$ in the bulk liquid (choices of historic origin [50, 51]), and the free energy of the small angle grain boundaries scales with parameter $H$ [144]. It is worth noting that due to the $[1 - p(\phi)]$ multiplier, the driving force of orientational ordering disappears in the bulk liquid. This ensures that a double counting of the orientational contribution to the free energy of the liquid is avoided, as this contribution has *per definitionem* been incorporated into the free energy density of the bulk liquid phase $f_L(c, T)$. As we are interested in solidification, which takes place on a considerably shorter time scale than grain boundary relaxation, the orientational mobility is assumed to vary proportionally to $p(\phi)$ across the interface, i.e., we set zero orientational mobility in the solid and the maximum value in the liquid. (This assumption can be relaxed, and grain boundary dynamics in the solid state can also be studied within the frame of the present model.) As a consequence of this assumption, orientational ordering takes place exclusively at the crystal-liquid interface, concurrently with structural ordering. An important consequence is that the orientational noise present in the inter-face region may contribute to the free energy of the solid-liquid interface. This can be avoided by an appropriate choice of the model parameters that leads to the development of an ordered liquid layer ahead of the solidification front (as observed in molecular dynamics simulations, see e.g. [76 − 78]). Under such conditions the orientational contribution to the interfacial free energy is insignificant, and the usual simple relationships between interfacial properties (thickness and free energy) and the model parameters remain valid.

With the introduction of the orientation field, *additional time* and *length scales* appear in the model. Specifically, the relaxation time of orientational perturbations is inversely proportional with the orientational mobility $M_\theta$, which in turn, is proportional to the *rotational* diffusion coefficient $M_\theta \propto D_{rot}$ of molecules that scales with the inverse viscosity (according to the Stokes-Einstein-Debye relationship) down to the glass transition temperature. Apparently, this new time scale plays a central role in the formation of polycrystalline structures. Recently, it has become appreciated that undercooled liquids of sufficiently high viscosity ($\approx 30$−$50$ Pa·s) exhibit spontaneous and long-lived heterogeneities, associated with the formation of regions within the fluid having much higher and lower mobility relative to a simple fluid in which particles exhibit Brownian motion [145, 146]. These *dynamic heterogeneities* persist on timescales of the order of the stress relaxation time, which can be minutes near the glass transition and astronomical times at lower temperatures. The presence of such transient heterogeneities has been associated with dramatic changes in the transport properties of supercooled liquids [147 - 151]. Specifically, both the translational diffusion coefficient $D_{tr}$ and the rotational diffusion coefficient $D_{rot}$ (quantities associated with the rate of molecular translation and rotation in the liquid) scale with the inverse of liquid shear viscosity at high $T$ and low undercooling, but $D_{rot}$ slows down significantly relative to $D_{tr}$ at lower $T$. This phenomenon in undercooled liquids is termed "decoupling" [147 - 151]. As a result, at low temperatures, where rotational relaxation is slow relative to the translational one that governs the growth rate, orientational defects (e.g. new grains) can be frozen into the solid. Model A naturally incorporates this possibility (the orientational mobility needs to be reduced relative to the phase field mobility). As it will be demonstrated in Sec. 2.5, Model C (a close relative of Model A) is able to recover many of the polycrystalline morphologies via the combination of this mechanism with polycrystalline branching of well-defined branching angle. Before reviewing these developments, we explore the applicability of the phase field theory for describing the formation of nanometer size heterophase fluctuations that initiate crystalline freezing.

## 2.4 Phase Field Theory of Crystal Nucleation

Crystallization of homogeneous non-equilibrium liquids is initiated by nucleation, during which crystallike *heterophase fluctuations* appear (Fig. 4) [152 - 155], whose formation is governed by the free energy gain when transferring molecules from liquid to the crystal and the extra free energy $\gamma$ needed to create the crystal-liquid interface. Those heterophase fluctuations that are larger than a critical size, determined by the interplay of the volumetric and interfacial contributions to the cluster free energy, reach macroscopic dimensions with a high probability, while the smaller ones dissolve with a high probability. Heterophase fluctuations of the critical size are termed *nuclei* and the process in which they form via internal fluctuations of the liquid phase is *homogeneous nucleation* (as opposed with the *heterogeneous nucleation*, where particles, foreign surfaces, or impurities help to produce the heterophase fluctuations that drive the system towards solidification). Even in simple liquids (such as Lennard-Jones), several local arrangements (bcc, fcc, hcp, icosahedral) compete [156, 157], of which often a metastable phase nucleates.

Before reviewing theory, it is appropriate to mention that recent experiments on colloidal suspensions and on biological systems shed much light on the microscopic aspects of crystal nucleation, and revealed phenomena that represent new challenges to theory. With appropriate surface treatments, the colloidal solutions mimic closely the hard-sphere liquid [158 - 162]. Since the size of the individual particles is in the micrometer range, crystallization can be monitored via light scattering. With modern experimental techniques (such as laser scanning confocal microscopy), real-time imaging of nucleation processes became possible [153]. This allows for a better characterization of the critical fluctuations, which have a quite irregular, fluctuating shape. The nucleation experiments on apofer-

(a)    (b)

(c)    (d)

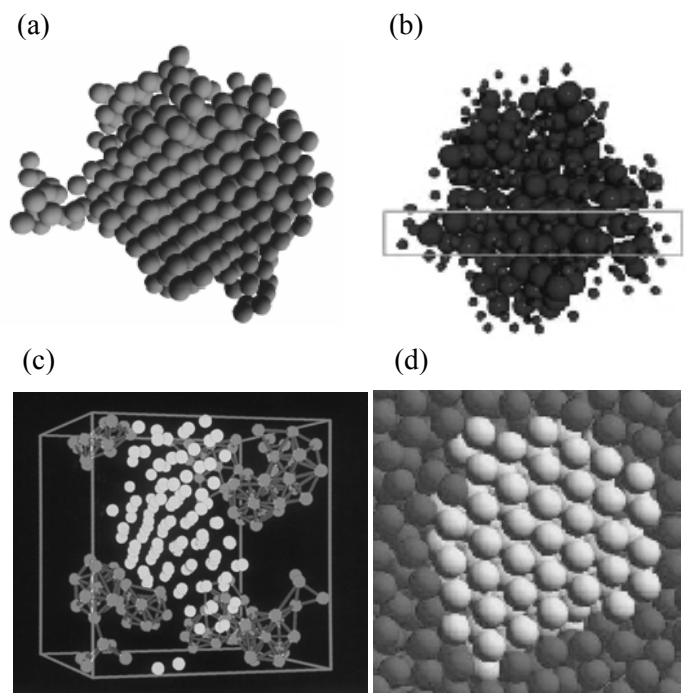

**Figure 4.** Crystalline heterophase fluctuations in non-equilibrium liquids. From left to right: (a) in Lennard-Jones liquid (molecular dynamics simulation, reprinted from [152], © 1996, with the permission of the American Institute of Physics); (b) in colloidal suspension (laser scanning confocal microscopy, reprinted from [153], © 2001, with the permission of the AAAS); (c) in Lennard-Jones glass (molecular dynamics simulation, reprinted from [154], © 1991, with the permission of Elsevier); (d) in hard-sphere liquid (molecular dynamics simulation, reprinted from [155], © 2001, with the permission of Nature Publishing Group).

ritin have shown that the critical fluctuations form via merging chainlike aggregates [163]. This unexpected behavior is thought to originate from a weak middle-range repulsion that precedes shorter-range attraction and core-repulsion between the molecules. This finding implies that nucleation is sensitive to details of the interaction potential, and it is essential to work out true molecular theories.

The description of the nanometer-size near-critical fluctuations is problematic even in one-component systems. The main difficulty is that critical fluctuations forming on reasonable experimental time scales contain typically a few times ten to several hundred molecules [152 − 173]. This together with the fact that the crystal-liquid interface extends to several molecular layers [32, 76 - 78] indicates that the critical fluctuations are essentially comprised of interface. Therefore, the droplet model of classical nucleation theory, which employs a sharp interface separating a liquid from a crystal with bulk properties, is certainly inappropriate for such fluctuations as demonstrated by recent atomistic simulations [155, 164].

Field theoretic models that predict a diffuse interface, offer a natural way to handle such difficulties [43, 44], and proved successful in addressing nucleation problems [167 − 173], including condensation [170, 171], and nucleation of metastable phases [172, 173]. Here, we review recent applications of the phase field theory for describing homogeneous crystal nucleation, and address two possibilities:

(a) The nucleation process can be *simulated* within the framework of the phase field theory. The proper statistical mechanical treatment of the nucleation process requires the introduction of uncorrelated Langevin-noise terms into the governing equations with amplitudes that are determined by the fluctuation-dissipation theorem [49, 54, 55, 65]. Such an approach has been used for describing homogeneous nucleation in a single-component [174] and binary systems [99, 100] and during eutectic solidification in a binary model [49, 65]. However, modeling of nucleation via Langevin-noise is often prohibitively time consuming. One remedy is simply to increase the amplitude of the noise. This, however, raises the possibility that the fluctuations, which initiate solidification, will most likely significantly differ from the real critical fluctuations. To avoid practical difficulties associated with modeling noise-induced nucleation, crystallization in simulations is often initiated by randomly placing supercritical particles into the simulation window [70, 71, 175, 176]. An alternative method has been proposed by Gránásy *et al.* [99, 100], who first calculate the properties of the critical fluctuations (see below) and then place such critical fluctuations randomly into the simulation window, while also adding Langevin-noise that decides whether these nuclei grow or dissolve.

(b) The phase field theory can also be used for *calculating the properties of the critical fluctuations and the height of the nucleation barrier* [99, 100, 177]. Being in unstable equilibrium, the critical fluctuation (the nucleus) represents an extremum of the free energy functional, subject to conservation constraints when the phase field is coupled to conserved fields. To mathematically impose such constraints one adds the volume integral of the conserved field times a Lagrange multiplier to the free energy. The field distributions, that extremize the free energy, obey the appropriate Euler-Lagrange (EL) equations, which in the case of the phase field theory take the form

$$\frac{\delta F}{\delta \phi} = \frac{\partial I}{\partial \phi} - \nabla \frac{\partial I}{\partial \nabla \phi} = 0 \ , \tag{4}$$

where $\delta F/\delta \phi$ stands for the first functional derivative of the free energy with respect to the field $\phi$, while $I$ is the total free energy density (that includes all the gradient terms). Here $\phi$ stands for all fields used in theory. The EL equations are solved under the appropriate boundary conditions: it is assumed that unperturbed liquid exists in the far field, while, for symmetry reasons zero field gradients prescribed at the center of the fluctuations. The same solutions can also be obtained as the non-trivial time-independent solution of the governing equations for field evolution. Having determined the solutions, the work of formation of the nucleus (height of the nucleation barrier) can be obtained by inserting the solution into the free energy functional.

As nucleation takes place at relatively large undercoolings, the interface thickness and the size of nuclei are comparable, and one can work with the physical interface thickness. Thus, one of the major obstacles of quantitative phase field modeling of large solidification objects forming at low undercoolings, i.e., the necessity to use unphysically broad interfaces, does not show up here. Furthermore, in the case of a few well-known model systems, all parameters of the phase field theory can be fixed, and the properties of the critical fluctuations can be calculated without adjustable parameters. For example, in the one-component limit of the standard binary phase field theory [50, 51], the free energy functional contains only two parameters, the coefficient of the square-gradient term for phase field and the free energy scale (height of the central hill between the double well in the local free energy density). If the thickness and the free energy of a crystal-liquid interface are known for the equilibrium crystal-liquid interface, all model parameters can be fixed and the properties of the critical fluctuation, including the height of the nucleation barrier, can be predicted without adjustable parameters. Such information is available from atomistic simulations/experiments for a few cases (Lennard-Jones system and ice-water system). This procedure leads to a good quantitative agreement with the magnitude of the nucleation barriers deduced from atomistic simulations for the Lennard-Jones system, and from experiments on ice nucleation in undercooled water [99, 100]. A similar approach for a binary Ni-Cu alloys lead to reasonable values for the temperature and composition dependence of the interface free energy of critical fluctuations, and also yielded reasonable critical undercoolings for electromagnetically levitated droplets [99, 100]. Similar results have been obtained for the hard-sphere system using a phase field model that relies on a structural order parameter coupled to the density field [178]. Again, the model parameters have been fixed via the interface thickness and interfacial free energy from atomistic simulations, so the calculations were performed without adjustable parameters. A similar approach has been used recently to address $CO_2$ hydrate nucleation in aqueous $CO_2$ solution under conditions typical to seabed hydrate reservoirs [179, 180].

Recent developments in atomistic modeling of small crystalline clusters in the hard-sphere system allowed for an extension of the analysis described in [178]. Cacciuto *et al.* [181] evaluated the free energy of clusters in the hard-sphere liquid of equilibrium density as function of size that allowed the determination of the size dependence of the solid-liquid interface free energy. The results extrapolate to $\gamma_{R \to \infty} = 0.616(3) \ kT/\sigma^2$, the cluster average of the interfacial free energy for infinite size ($\sigma$ is diameter of the hard spheres). This value agrees well with results from molecular dynamics simula-tions (e.g., with $\bar{\gamma}/(kT/\sigma^2) = 0.612 \pm 0.02$ for the average of the values for the (111), (110), and (100) directions by Davidchack and Laird [89]; and with $\bar{\gamma}/(kT/\sigma^2) = 0.63 \pm 0.02$ by Mu *et al.* [182]). This allows the fixing of the coefficient of the square-gradient term with a higher accuracy than in previous work, since it was uncertain how far the cluster (or orientational) average of the interfacial free energy falls from the average for the (111), (110), and (100) directions. A further refinement of the theory is that the density dependence of the coefficient of the square-gradient term, $\varepsilon^2 \propto C''(k)$, and of the free energy scale, $w \propto 1/S(k)$, were taken into consideration, where $C(k)$ is the direct correlation function of the liquid, related to the structure factor of the liquid as $S(k) = 1/[1 - C(k)]$, and $C''$ is its second derivative with respect to its argument. The parameter-free predictions of the PFT and the exact Monte Carlo results are compared in Fig. 5 [183]. The agreement between theory and MC simulations is convincing; considerably better than the (parameter-free) predictions of the classical nucleation theory and the self-consistent classical theory of Girshick and Chiu [184], while it is somewhat better than the parameter-free prediction by the phenomenological diffuse interface theory of Gránásy [185]. The uncertainty of the input data (interfacial free energy, equations of state, etc.) does not influence this result perceptibly [183].

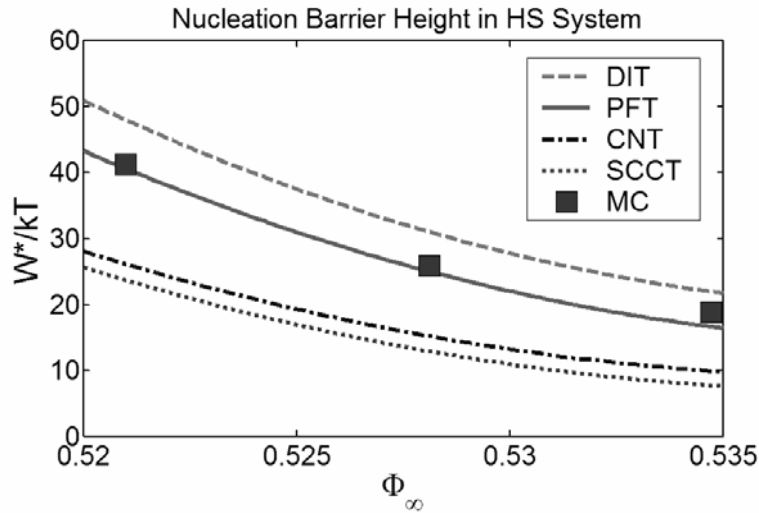

**Figure 5.** The height of the nucleation barrier vs. the initial density of the hard-sphere liquid as predicted by the phase field theory (PFT [178]), the sharp interface droplet model of the classical nucleation theory (CNT); the self-consistent classical theory (SCCT [184]), and the phenomenological diffuse interface theory (DIT [185]). These calculations do not contain adjustable parameters. For comparison the height of the nucleation barrier from Monte Carlo simulations (MC [155, 164]) is also presented.

These findings suggest that, using the physical interface thickness, the phase field theory is able to predict the height of the nucleation barrier quantitatively. This success [99, 100, 178, 183], together with the parameter-free prediction of the dendritic growth rate [32, 58], suggests that a multi-scale approach of the phase field theory with model parameters deduced from atomistic simulations is capable for quantitative predictions for both crystal nucleation and growth.

Below we describe further advances in the theory of polycrystalline solidification, particularly in the directions of restoring the rotational invariance of the free energy and incorporating a natural (noise-driven) nucleation of new crystal orientations.

## 2.5. Crystallization Kinetics

### 2.5.1. Polycrystalline Solidification in the Ideal Solution Model

Model A has been used to study the kinetics of anisotropic multi-particle solidification in two dimensions in binary ideal solutions (Ni-Cu) [99, 100]. A polycrystalline dendritic morphology closely resembling Fig. 1(b) has been obtained (see Fig. 6). The large number of particles (722) provides reasonable statistics for evaluating the Avrami-Kolmogorov exponent $p$. Transformation kinetics emerging from four representative simulations performed on 7000×7000 grids are compared [101]: Two of the simulations were performed for the same normalized initial liquid concentration of $x = (c_\infty$

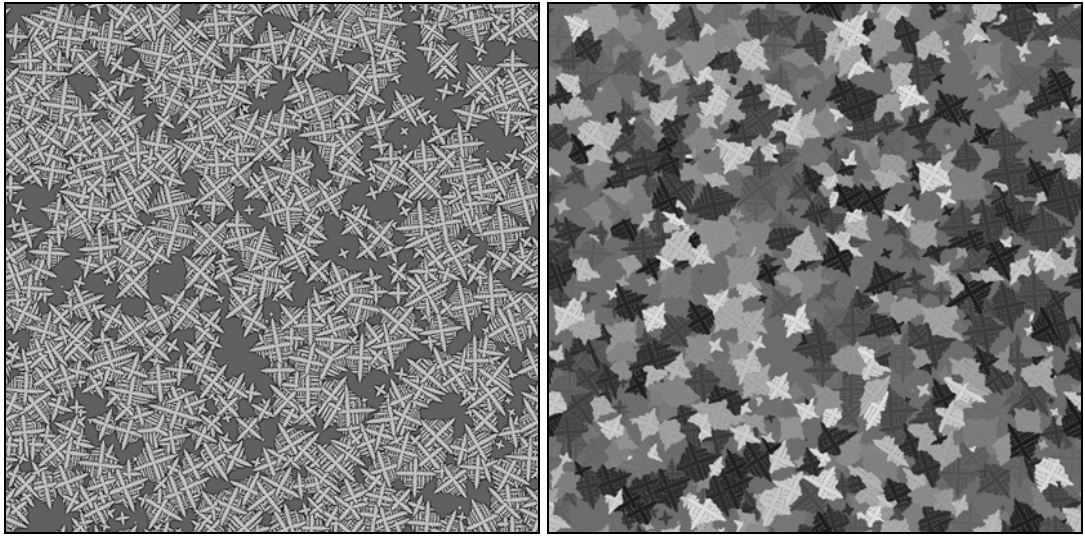

**Figure 6.** Snapshots of the concentration (left) and orientation (right) fields for two-dimensional dendritic solidification of a binary alloy (Ni-Cu) as predicted by Model A at 1574 K and supersaturation 0.8. By the end of solidification 722 dendritic particles formed. The calculation has been performed on a 10,000 × 10,000 grid (131.3 μm × 131.3 μm) with a 10% anisotropy of the interfacial free energy that was assumed to have a four-fold symmetry. (On the left, black and white correspond to the solidus and liquidus compositions, respectively, while the intermediate compositions are shown by hues that interpolate linearly between these colors. On the right, different hues denote different crystallographic orientations. When the fast growth direction is upward, 30, or 60 degrees left, the grains are shaded dark, light, or middle gray, respectively, while the intermediate angles are denoted by a continuous transition among these hues. Owing to the four-fold symmetry, orientations that differ by 90 degree multiples are equivalent.)

$- c_s)/(c_l - c_s) = 0.2$, close to the solidus composition ($c_\infty$ is the composition of the initial liquid, $c_s = 0.399112$ and $c_l = 0.466219$ are the solidus and liquidus compositions at $T = 1574$ K). The other two simulations were performed at higher solute contents ($x = 0.5$ and 0.8), that lie midway between the solidius and liquidus compositions, and close to the liquidus, respectively. Representative 1000×1000 sections of these simulations are shown in Fig. 7 [panels (a) – (d)], together with the respective

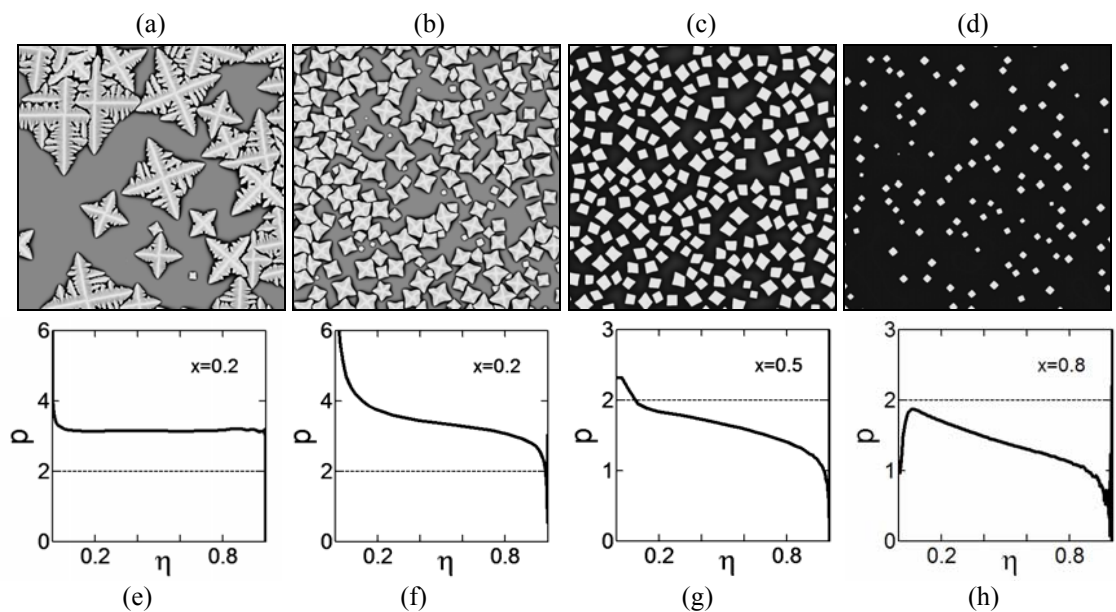

**Figure 7.** Two-dimensional anisotropic multigrain solidification as a function of composition and nucleation rate in the Cu-Ni system at 1574 K as predicted by Model A. (a)-(d) 1000×1000 segments (13.2 μm × 13.2 μm) of the concentration distribution (white: solidus; black: liquidus); (e)-(h) the respective Avrami-Kolmogorov exponent vs. normalized transformed fraction curves are shown. Simulations presented in panels (a) and (b) differ in the magnitude of the nucleation rate.

Avrami-Komogorov exponents evaluated as a function of the normalized crystalline fraction $\eta = X/X_{max}$, [panels (e) − (h)], where $X_{max}$ is the maximum crystalline fraction achieved at the given liquid composition. Note the morphological transition from the dendritic structure towards the equilibrium shape with decreasing supersaturation (defined as $S = 1 − x$).

In the simulation shown in Fig. 7(a) the nucleation rate is sufficiently low to enable the formation of fully developed dendritic structures. It is worth noting in this respect that in the case of dendritic solidification, the global average of the composition of the growing solid combined with the interdendritic liquid trapped between the dendrite arms must be equal to the initial composition of the liquid (as required by mass balance), thus solute pile up does not decelerate the advance of the perimeter (except as a transient), which is determined essentially by the growth velocity of the dendrite tips. Since the dendrite tip is a steady state solution of the diffusion equation, a constant nucleation and growth rates apply here, and $p = 1 + d = 3$ is expected for the Avrami-Kolmogorov exponent in two dimensions. The observed value, $p \approx 3$, is fully consistent with this expectation. In the other three simulations, the particles have more compact shapes, and interact via their diffusion fields, a phenomenon termed 'soft impingement'. The respective Avrami-Kolmogorov exponents decrease with increasing solid fraction. A closer inspection of the process indicates that at large supersaturations where there is no substantial compositional difference between the nucleus and the initial liquid [see Figs. 7(b) and 7(f)], supercritical growth right after nucleation is governed by the phase field mobility, as opposed to chemical diffusion controlled growth at later stages. This transient period represents a delay in the onset of diffusion-controlled growth, resulting in a value for $p$ that decreases with time; a phenomenon that becomes weaker with decreasing supersaturation. This phenomenon is expected to be perceptible in only the case of copious nucleation, where the length of the transient period is comparable to the total solidification time. Such behavior has been indeed observed experimentally during the formation of nanocrystalline materials made via the devitrification of metallic glass ribbons [127], a process characterized by enormous nucleation rates.

### 2.5.2. Polycrystalline Solidification in the Regular Solution Model

Pusztai and Gránásy introduced regular solution thermodynamics into Model A, which can be then used to describe simple eutectic and peritectic systems [186]. It is worth recalling that in many eutectic systems, the two solid phases have a well-defined orientational relationship [187 − 189]. To address such a situation, an extra free energy term has been incorporated that prefers a fixed misorientation at the phase boundaries [186]. Since the model contains a single structural order parameter (phase field), it is strictly applicable only to systems, where the two phases have the same crystal structure (e.g., Ag-Cu, Ag-Pt).

(a)                                    (b)                                    (c)

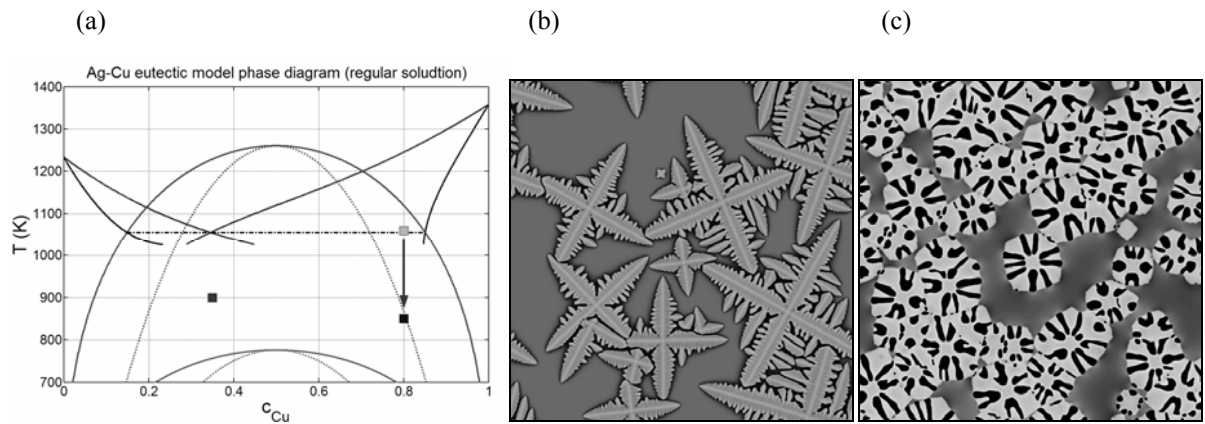

**Figure 8.** Solidification in the Ag-Cu system as predicted by the phase field theory: (a) Regular solution phase diagram for the Ag-Cu system. Solidification morphologies in the gray point and the black point on the right in panel (a), respectively: (b) Primary dendritic solidification. (c) Equiaxed eutectic grains formed by noise-induced nucleation. Composition maps are shown.

Two-dimensional simulations have been performed for the Ag-Cu system [Fig. 8(a)] at various initial liquid compositions (hypo-eutectic, eutectic, and hyper-eutectic), which showed that the model successfully accounts for the orientational locking of the solid phases (Fig. 9). the Avrami-Kolmogorov exponent $p$ has also been evaluated. In agreement with experiment [190] and phase field simulations without orientation field [49, 65], the value $p \approx 3$ obtained follows the $p = 1 + d$ rule, where $d$ is the number of dimensions (Fig. 10).

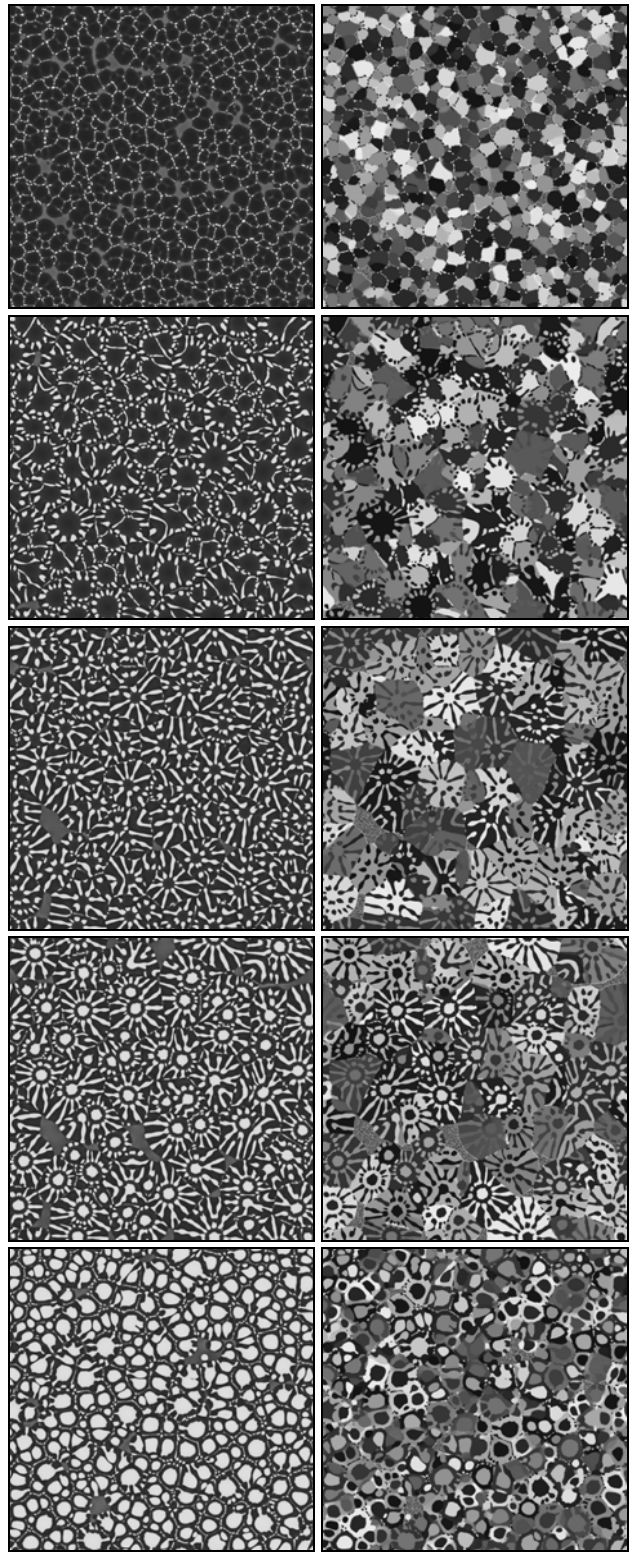

**Figure 9.** Equiaxed solidification in hypo-eutectic (from top to bottom: $c_{Cu}$ = 0.2, 0.3), eutectic ($c_{Cu}$ = 0.35), and hyper-eutectic ($c_{Cu}$ = 0.4, 0.5) Ag-Cu liquids at 900 K as predicted by the phase field theory. Composition maps are shown in the top row, the respective orientation maps are in the bottom row. (In the composition maps, continuous change from black to white indicates compositions varying from $c_{Cu}$ = 0 to 1, respectively. In the orientation maps, different hues stand for different crystallographic orientations in the laboratory frame.) Note the locked (fixed) misorientation of the two phases within the eutectic particles.

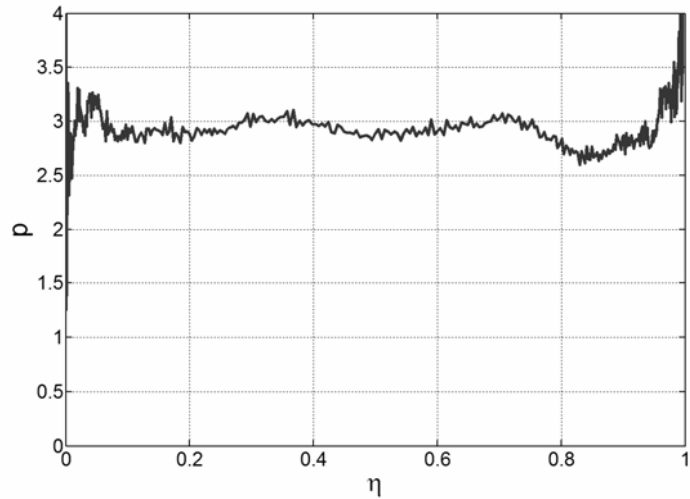

**Figure 10.** Avrami-Kolmogorov exponent for eutectic solidification as a function of transformed fraction, determined in the point marked by the rightmost black square in Fig. 8(a).

The dynamic recovery of natural lamellar spacing via initiating eutectic solidification with significantly lower and larger spacing than the natural one has also been explored using this model. Our simulations imply that the natural lamella spacing is re-established via the nucleation of new lamellae (Fig. 11). To investigate morphologies associated with primary dendritic solidification and subsequent eutectic solidification, two-stage heat treatments were performed (above and below the eutectic temperature). The results are presented in Fig. 12. The solidification started with epitaxial growth of the primary phase, followed by copious-nucleation-driven eutectic solidification.

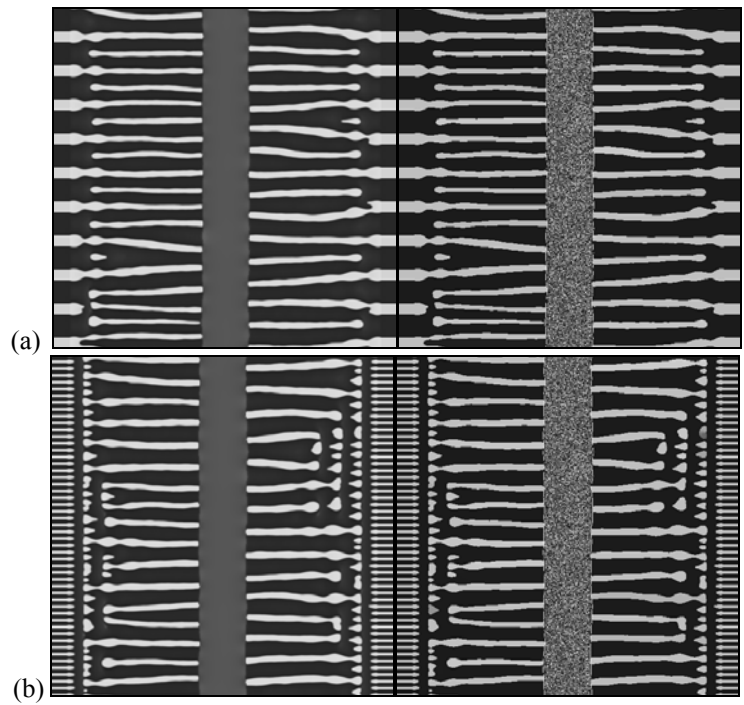

**Figure 11.** Dynamic self-adjustment of eutectic lamellar spacing: (a) Initial wavelength is larger than the natural one. (b) Initial wavelength is smaller than the natural wavelength. Composition (left) and orientation maps (right) are shown.

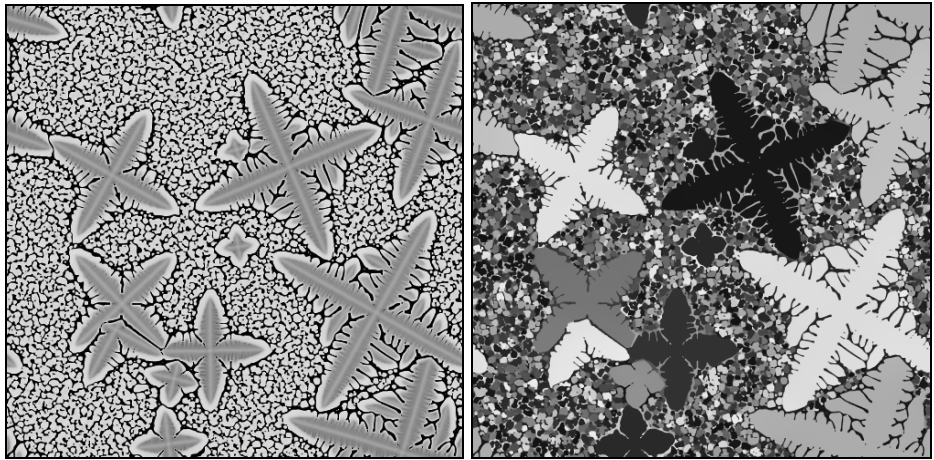

**Figure 12.** Solidification morphology after primary dendritic crystallization took place above the eutectic temperature (gray square in Fig. 8[a]), and subsequent heat treatment below the eutectic temperature (black square on the right in Fig. 8[a]).

The same model has been used to describe peritectic solidification in a regular solution model system with interaction parameters chosen so that the qualitative features of the Ag-Pt phase diagram are recovered (Fig. 13). The phase field simulations have been performed inside the metastable liquid spinodal (red square inside the region bounded by dashed green line in Fig. 13). Accordingly, solidification is preceded by liquid phase separation. As found in molecular dynamics [191] and density functional [192] studies for other systems, this process helps crystal nucleation, which takes place in the high Ag-content liquid droplets (Fig. 14).

In order to enable the modeling of the formation of a metastable crystalline phase besides the stable one, an extra solid-solid structural order parameter $\psi$ has been introduced (see Model B in Appendix). Combined epitaxial and equiaxed solidification at the eutectic composition in the Ag-Cu system is shown in Fig. 15. Initially solid layers were placed to the left and right sides of the simulation box (low Cu content solid of structure stable at this composition). Note the correlation of the solid-solid order parameter, the composition, and the orientation field, and the *transition from island growth to lamellar growth* in the expitaxial regions on the left and right sides of the simulation window. Work is underway to explore competing nucleation and growth of metastable and stable phases.

These results indicate that Models A and B reproduce the essential features shown by eutectic / peritectic solidification.

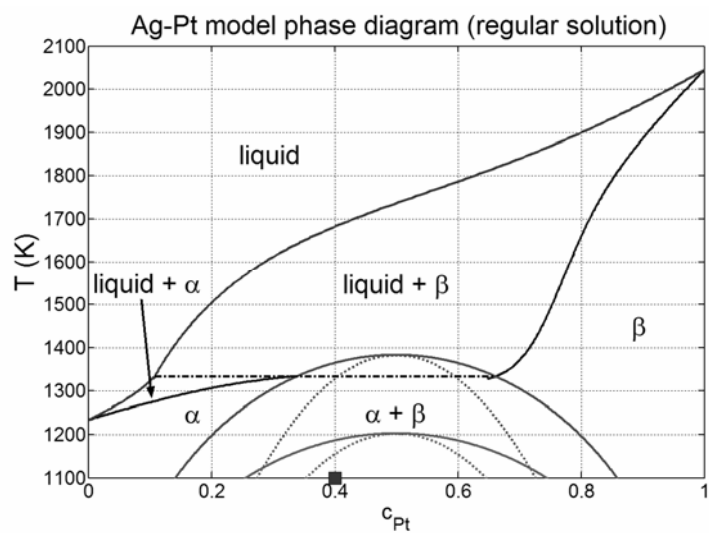

**Figure 13.** Phase diagram of model peritectic system (regular solution Ag-Pt). The parameters of the regular solution model have been chosen so that the general features of the experimental phase diagram are reproduced.

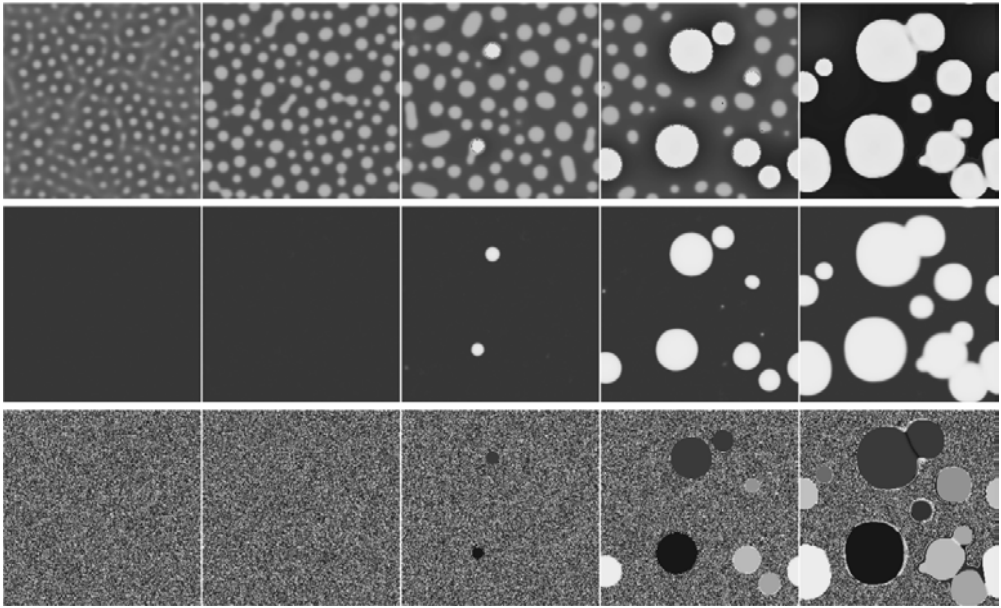

**Figure 14.** Solidification in a model peritectic system shown in Fig. 13 as predicted by the phase field theory. The transformation starts with liquid phase separation assisted nucleation of the crystals. Upper row: composition map (black: pure Ag, white: pure Pt); central row: phase field, bottom row: orientation field.

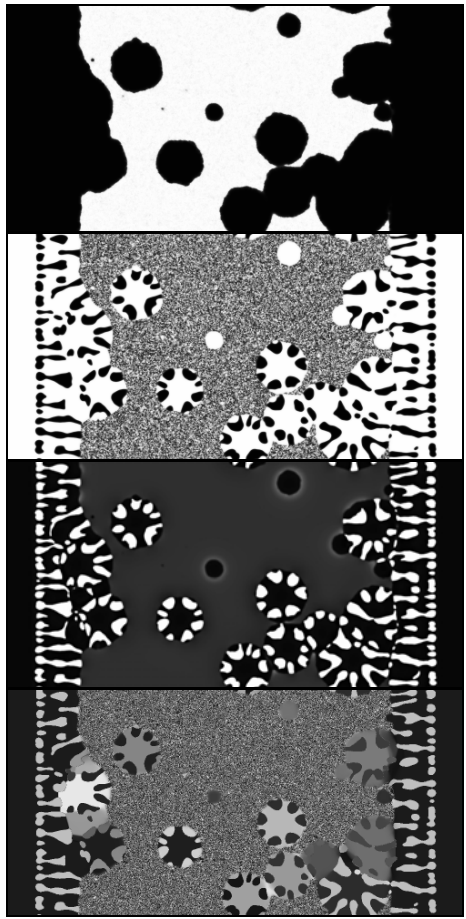

**Figure 15.** Epitaxial and equiaxed growth during eutectic solidification as predicted by the four-field phase field theory (Model B, $T$ = 900 K, $c_{Cu}$ = 0.35). From top to bottom: Solid-liquid phase field, solid-solid phase field, composition, and orientation fields. Note the correlation of the latter fields, and the fixed orientational relationship between the two solid phases.

## 2.6. Polycrystalline Growth Forms

Particulate additives are known to influence the solidification microstructure. Such additives are used as grain refiners for many practical systems as they lower the nucleation barrier for the crystalline phase and can thus be used to control the number density of crystalline particles that form during the solidification process. Recent experiments on clay filled polymer blend films revealed that, besides this role, particulate additives may also perturb crystal growth, yielding polycrystalline growth morphologies [6]. Remarkably, polycrystalline growth also occurs in pure liquids in the absence of particulate additives (e.g., [7, 193, 194]). Both routes to *polycrystalline growth* have recently been addressed within the framework of the phase field theory.

### 2.6.1. Particle-Induced Grain Nucleation at the Perimeter

A spectacular class of structures appears in thin polymer blend films if foreign (clay) particles are introduced [6, 195]. This disordered dendritic structure is termed a 'dizzy' dendrite [Fig. 1(c)] and form by the engulfment of the clay particles into the crystal, inducing the formation of new grains. This phenomenon is driven by the impetus to reduce the crystallographic misfit along the perimeter of clay particles by creating grain boundaries within the polymer crystal. This process changes the crystal orientation at the dendrite tip, changing thus the tip trajectory ('tip deflection'). To describe this phenomenon, Gránásy *et al.* [195] incorporated a simple model of foreign crystalline particles into Model A: They are represented by *orientation pinning centers*, — small areas of random, but fixed orientation — which are assumed to be of a foreign material, and not the solid $\phi = 0$ phase. This picture economically describes morphological changes deriving from particle-dendrite interactions.

The simulations (see Fig. 16) show that tip deflection occurs only when the pinning center is above a critical size, comparable to the dendrite tip radius. Larger pinning centers cause larger deflections. With increasing orientational misfit between the particle and the dendrite, dendrite tip deflection was found to increase. However, above a critical angular difference between the pinning center and the dendrite ($\Delta\theta \approx 0.35$), the pinning center is simply engulfed into the dendrite without deflection, while the tip splits to some extent. This is due to the high interface energy at these misorientations, creating an energetic preference for a small layer of liquid around the inclusion. In this case the wet phase boundary appears as a hole in the crystal. An important consequence of this effect is that the angle of tip deflection has an upper limit, thus preventing large deviations from the original growth direction. Pinning centers cause deflection only if *directly* hit by the dendrite tip, a finding confirmed by experiment. This explains the experimental observation that only a small fraction of the pinning centers influence morphology. Using an appropriate density of pinning centres comparable to the density of clay particles, a striking similarity is obtained between experiment and simulation (Fig. 17). This extends to such details as curling of the main arms and the appearance of extra arms. The disorder in dendrite morphology originates from a polycrystalline structure that develops during a sequential deflection of dendrite tips on foreign particles.

Recent studies by Gránásy *et al.* [33, 196] show that increasing the number density of the foreign particles (orientation pinning centers) further increases the 'randomness' of the solidification morphology leading to a continuous transition from a 'dizzy' dendrite into the *seaweed* morphology (see discussion in Sec. 2.6.3). The perturbation caused by the foreign particle at the growth front leads to the formation of new crystal grains at the perimeter of the growing particle. Therefore, it has been termed as foreign-particle-induced growth front nucleation (GFN).

Gránásy *et al.* investigated possible control of growth morphologies combining this mechanism with 'intelligent' orientation pinning centers [195]. The effect of uniformly oriented orientation pinning centers of fixed orientation is shown in Fig. 18. The dendrite arms bend so that after a brief transient their fast growth directions coincide with those dictated by the pinning centers. In contrast, the uniformly rotating orientation pinning centers yield spiraling dendrite arms and a periodic concentric ring structure for the orientation field (Fig. 19). Parallel pinning lines of alternating orientation lead to zigzagging dendrite arms and a striped orientation map (Fig. 20). Experimental realization of these complex pinning conditions is certainly a challenge. Possible control methods might include the use of substrate-embedded oriented particles, the rotation of an external electromagnetic field, or angular momentum control by laser pulses [197]. Previous work has shown that particles are not neces-sary to nucleate crystallization in the thin polymer films, and that nucleation can be achieved by simply piercing the film with a sharp glass fiber [198]. By extension, it should be possible to print arrays of nucleation sites with specified symmetry of configuration by simply rolling a cylinder with an array of asperities over the uncrystallized polymer film. The orientation of nucleation points could be controlled by making the asperities in the form of flat pins with controlled orientation. In this way, it should be possible to create a wide range of crystallization morphologies and to tune the topography, permeability and the mechanical properties of the crystallized polymer film. Such orientation-controlling techniques may open a new route for tailoring solidification microstructures.

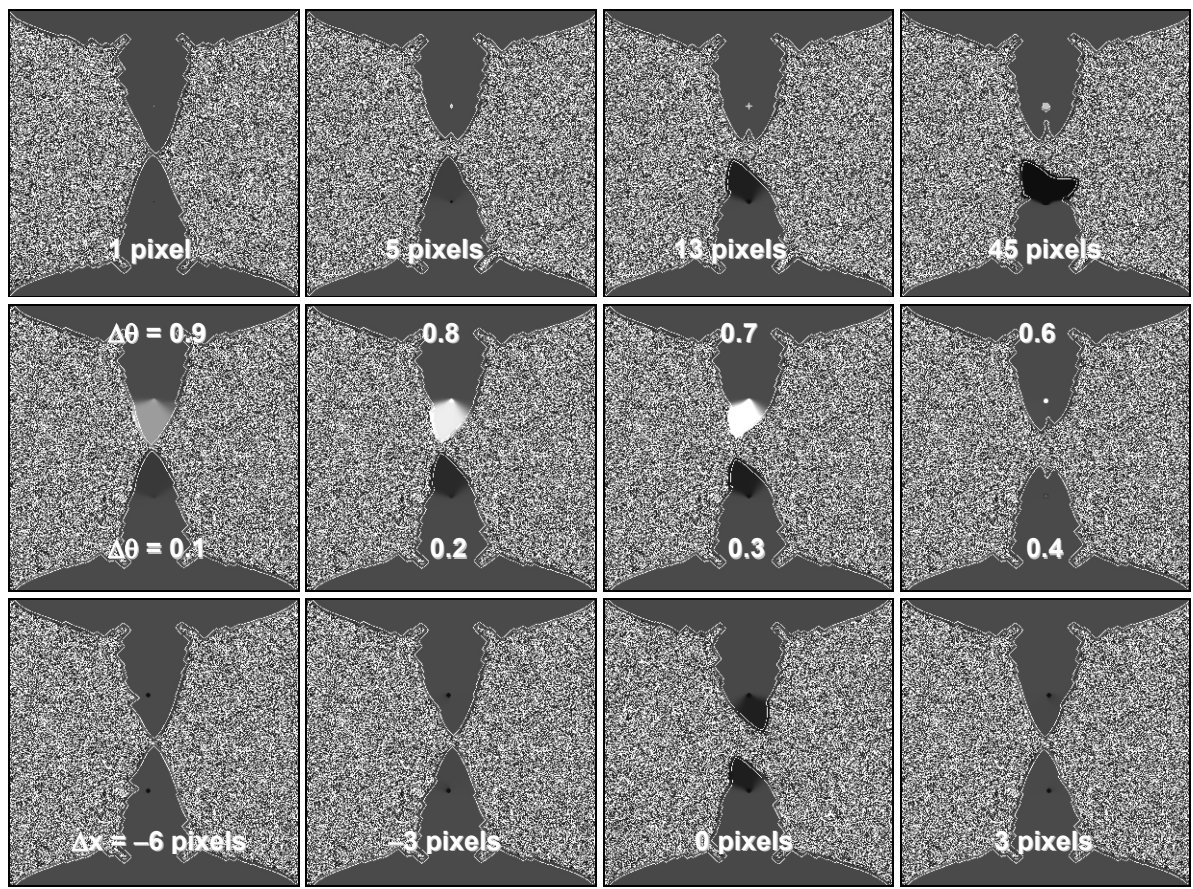

**Figure 16.** Deflection of a dendrite tip by an orientation pinning center in Model A. The first row shows the influence of the size of the pinning center: larger pinning centers cause larger deflections (the misorientation $\Delta\theta$ is set to 0.333 below and 0.5 above). In the middle row the effect of increasing misorientation $\Delta\theta$ of (13 pixel-sized) pinning centers is shown. As the angle increases beyond 0.3 (or less than 0.7 by symmetry) the effective surface energy increases to the point where the boundary prefers to be 'wet', which results in tip splitting as opposed to deflection. The third row shows that unless the tip is precisely lined up with the (13 pixel) pinning center, the tip does not deflect, even though the misorientation is $\Delta\theta = 0.3$. $\Delta x$ is the lateral disposition. (Grayscale is the same as for the right panel of Fig. 6.) The simulations were performed on a 300×300 rectangular grid (4 μm × 4 μm), with the thermodynamic properties of Ni-Cu, and 15 % anisotropy of the interface free energy. $\theta$ is normalized to vary between 0 and 1.

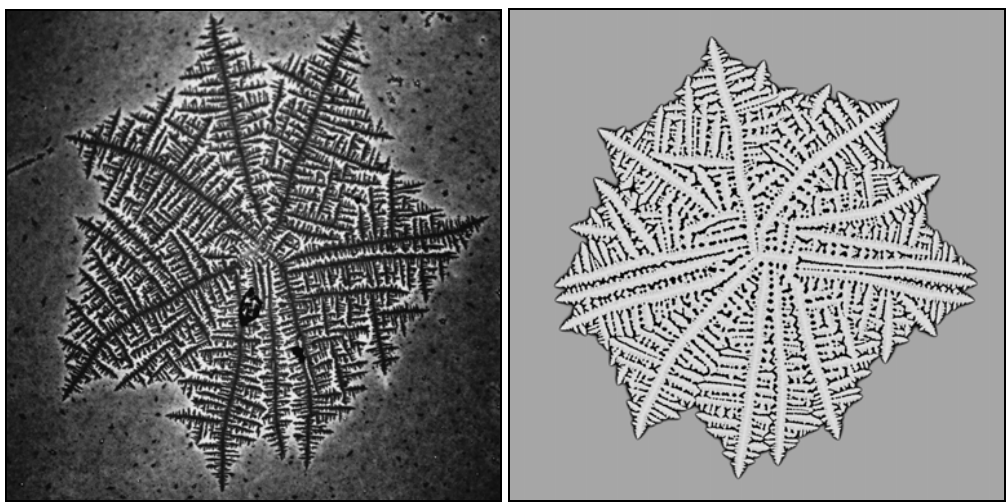

**Figure 17.** Disordered ("dizzy") dendrites formed by sequential deflection of dendrite tips on foreign particles: Comparison of experiments on 80 nm clay-polymer blend film (left, courtesy of V Ferreiro and J F Douglas; for the experimental details see [6]) and phase field simulation by Model A (right). (The simulation was performed on a 3,000 × 3,000 grid (39.4 μm × 39.4 μm), with 18,000 single-pixel orientation pinning centers per frame.)

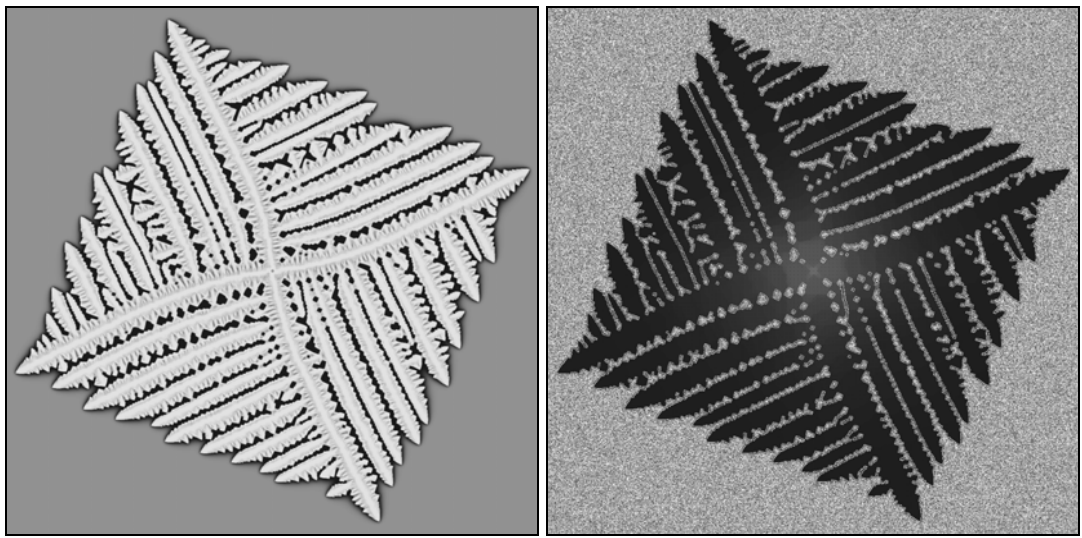

**Figure 18.** The effect of uniformly oriented foreign particles (black: orientation with the fast growth direction 30 degrees left of vertical) on a dendrite nucleated with the fast growth direction upwards (middle grey tone) in Model A. (Left: composition map; right: orientation map. The simulations were performed on a 3000 × 3000 grid (39.4 μm × 39.4 μm). Grayscale is the same as for the right panel of Fig. 6.)

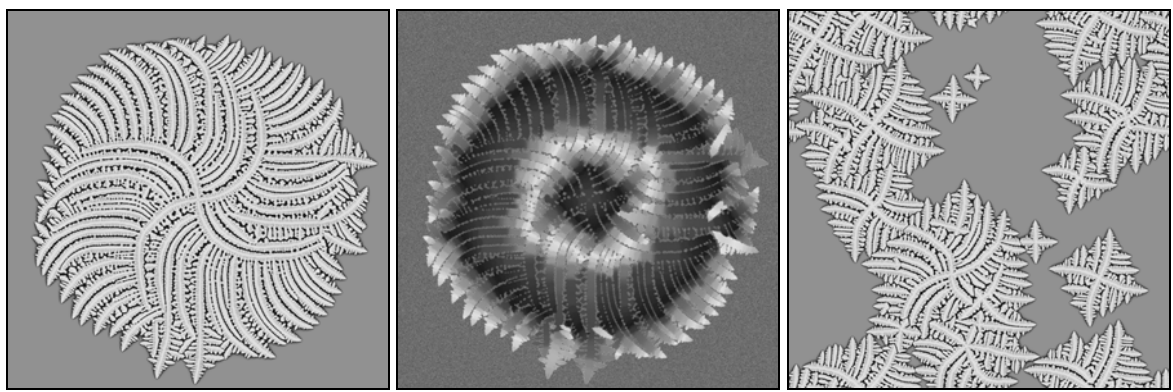

**Figure 19.** The effect of uniformly rotating foreign particles on a dendrite nucleated with the fast growth direction upwards (middle gray tone) in Model A. Note the spiraling arms and the periodic concentric ring structure of the local crystal orientation that preserves the temporary orientation at solidification. (Left: concentration map; center: orientation map; right: concentration map for a simulation with nucleation switched in.) The simulations were performed on a 3000 × 3000 grid (39.4 μm × 39.4 μm). Grayscale is the same as for the right panel of Fig. 6

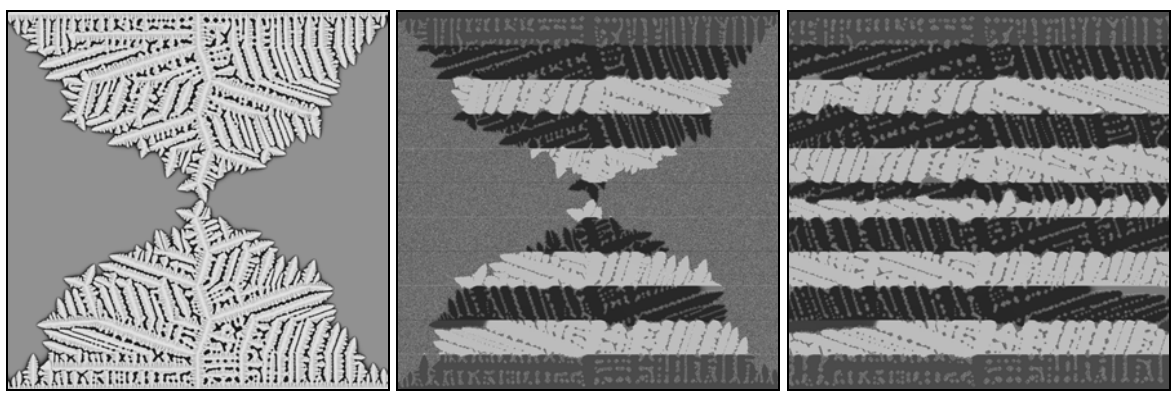

Figure 20. The effect of orientation pinning lines (dark and light gray orientations – corresponding to having the fast growth direction 30 and 60 degrees left from vertical) on a dendrite nucleated with the fast growth direction upwards (middle gray tone) in Model A. (Left: composition map; right: orientation map.) Note the striped orientation structure. (Left: concentration map; center: orientation map; right: late stage of concentration map.) The simulation was performed on a 1,800 × 1,800 grid (23.6 μm × 23.6 μm). Grayscale is the same as for the right panel of Fig. 6.)

### 2.6.2. Polycrystalline Growth in the Absence of Foreign Particles

The mechanism described above is certainly not a general explanation for polycrystalline growth since spherulites have been observed to grow in liquids without particulates or detectable molecular impurities. How can this be understood? A clue to this phenomenon can be found in the observations of Magill [194], who noted that spherulites only seem to appear in highly undercooled pure fluids of sufficiently large viscosity. Interpreting Magill's observations, Gránásy *et al.* hypothesized [33, 196] that the decoupling of the translational and rotational diffusion coefficient at low temperatures is responsible for the propensity for polycrystalline growth in highly undercooled liquids. Specifically, a reduced $D_{rot}$ makes it difficult for the newly forming crystal regions to reorient with the parent crystal at the growth front advancing with a velocity that scales with the translational diffusion coefficient. Thus epitaxy cannot keep pace with solidification, i.e., the orientational order that freezes into the solid is incomplete. This situation can be captured within the phase field theory by reducing the orientational mobility while keeping the phase field mobility constant as discussed in detail by Gránásy *et al.* [33, 101, 196].

The first step in this direction has been made by Gránásy *et al.* [101], who reported the formation of polycrystalline spherulite in Model A, when reducing the orientational mobility at large driving force. In previous calculations for the impingement of single crystals, the orientational mobility was set so that single crystal nuclei formed (see Fig. 21, left panel). (Though occasionally, a single nucleation event could initiate the simultaneous appearance of several orientations in the simulations – i.e., the formation of polycrystalline nuclei has been observed, a phenomenon observed also in experiments [199] and atomistic simulations [200].) When the orientational mobility is significantly reduced, the system cannot establish the same orientation along the perimeter of the particle; only local orientational ordering may take place, leading to the formation of new crystal grains with different crystal orientations (Fig. 21, right panel). This process, which is responsible for the polycrystalline growth forms in pure systems, leads to similar results as the particulate heterogeneities, and represents a second form of growth front nucleation.

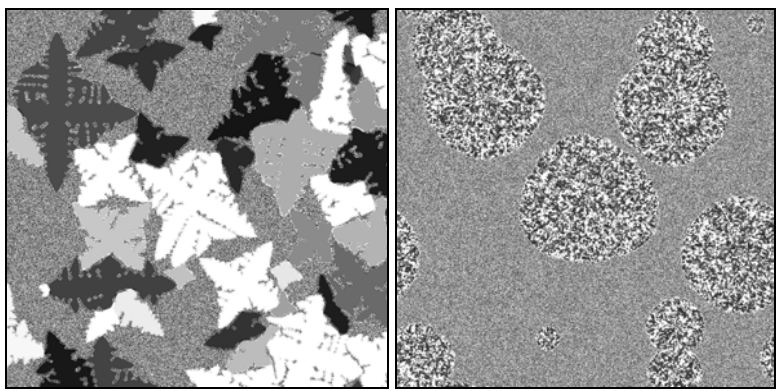

**Figure 21.** The effect of reduced orientational mobility of solidification morphology (Model A): Left: The orientational mobility set so that a single orientation emerges from the nucleation events (single crystal nuclei). Right: Polycrystalline structure forming when the orientational mobility is reduced by a factor of 1/30. The system cannot establish the same orientation along the perimeter; only local ordering is possible, yielding roughly spherical polycrystalline objects consisting of a large number of fine grains. Note that anisotropy is averaged out due to the large number of randomly oriented fine grains. (Orientation maps are shown. Grayscale is the same as for the right panel of Fig. 6. The simulation was performed on a $1,500 \times 1,500$ grid (19.7 µm × 19.7 µm).)

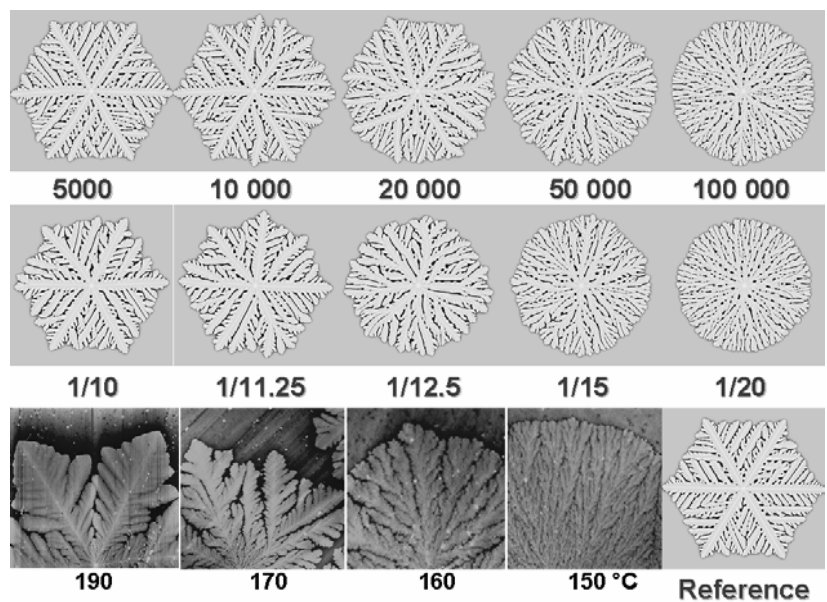

**Figure 22.** The duality of the effects of foreign particles (upper row) and reduced orientational mobility (central row) on the *polycrystalline* growth morphology (Model A). The number of the foreign particles per frame and the factor by which the orientation mobility was reduced are given below the panels. Composition maps are shown (white: solidus; black: liquidus). The simulations were performed on a 1000 × 1000 grid (13.1 μm × 13.1 μm). For comparison experimental images showing a similar morphological transition on polymer films is also shown (lower row; experimental details are given in [203]), together with the reference (single crystal) dendritic structure (lower row, rightmost panel).

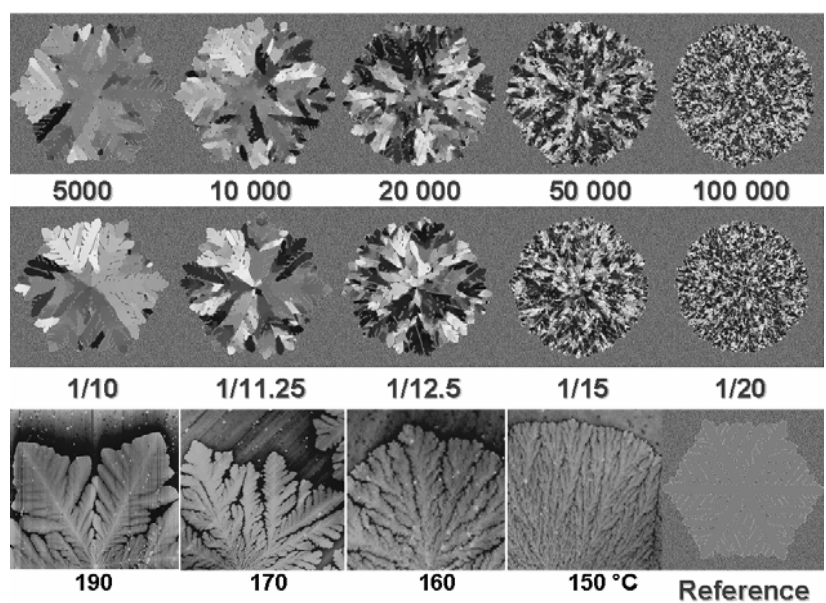

**Figure 23.** Orientation maps for the images shown in Fig. 22, where available (no orientation map is available for the experimental images). (Grayscale is the same as for the right panel of Fig. 6.)

### 2.6.3. Duality of Static and Dynamic Heterogeneities

Gránásy *et al.* [33, 196] recently performed systematic studies of polycrystalline morphologies forming via particulate induced GFN and low orientation mobility induced GFN. They observed that the two mechanisms lead to strikingly similar morphologies and grain structures (see Figs. 22 and 23). These results demonstrate a duality between the morphologies evolving due to the effects of static heterogeneities (foreign particles) and dynamic heterogeneities (quenched-in orientational defects).

It is worth noting in this respect that a dendrite to polycrystalline seaweed transition has been observed in electrodeposition [201], and that polycrystalline seaweed structures are commonly observed in electrochemical processes [12] or during the crystallization of electrodeposited layers [201, 202]. Similar morphological transition has been seen during the crystallization of thin polymer films [203]. Despite the success of modeling fractal-like morphologies on the basis of diffusion-limited-aggregation [204, 205], details of the polycrystalline seaweed formation are poorly understood. Quenching of orientational defects into the crystal due to reduced rotational diffusivity under coupling with diffusion controlled fingering (as happens in our phase field model), offers a straightforward explanation for both the morphology and the polycrystalline nature.

### 2.6.4. Formation of Spherulites

Spherulites are ubiquitous in solids formed under highly non-equilibrium conditions [193]. As mentioned in the introduction, they are observed in a wide range of metallurgical alloys, in pure Se [7], in metallic and oxide glasses [190, 206], mineral aggregates and volcanic rocks [207, 208], polymers [194, 209], liquid crystals [210, 211], simple organic liquids [212], and diverse biological molecules [3, 213-215]. Many everyday materials, ranging from plastic grocery bags to airplane wings and cast iron supporting beams for highway bridges, are fabricated by freezing liquids into polycrystalline solids containing these structures. The properties and failure characteristics of these materials depend strongly on their microstructure, but the factors that determine this microstructure remain poorly understood.

While the term 'spherulite' suggests a nearly spherical shape (circular shape in two dimensions where the term spherulite is still employed), this term is used in a broader sense of densely branched, polycrystalline solidification patterns [8 – 11, 193, 194, 207 – 225].

Experimental studies performed over the last century indicate that there are two main categories of spherulites [11, 216, 217]. Category 1 spherulites grow radially from the nucleation site, branching intermittently to maintain a space filling character (Fig. 24). In contrast, Category 2 spherulites grow initially as thread-like fibers, subsequently forming new and new branches at the growth front (Fig. 24). This branching of the crystallization pattern ultimately leads to a crystal 'sheaf' that increasingly splays out during growth. At still longer times, these sheaves develop two 'eyes' (uncrystallized regions) on each side of the primary nucleation site. Ultimately, this type of spherulite settles down into a spherical growth pattern, with eye structures apparent in its core region. In some materials, both categories of spherulite occur in the same material under the same nominal thermodynamic conditions.

The formation of spherulites has been addressed by various theoretical approaches that relate the large scale structure to the diffusion length [193] or to the wavelength of the Mullins-Sekerka instability [226]. Nucleation controlled growth of polymers has also been explored [227, 228]. One of the popular ideas used to explain the formation of spherulites envisions a regular branching of crystalline filaments with well-defined branching angle (see e.g., [7, 11, 194, 218]). While the details of such a mechanism necessarily differ on the molecular scale for the many systems that display spherulitic solidification, we hope to capture the general features of this process. To incorporate *branching with a fixed orientational misfit*, we included a new form of the orientational free energy (see Model C in Appendix). Here the orientational free energy has a second (local) minimum as a function of misorientation angle $\xi_0 |\nabla\theta|$, where $\xi_0$ is the correlation length of the orientation field. Thus, during orientational ordering at the solid-liquid interface, a second low free energy choice (preferred misorientation) is offered. Accordingly, the cells that have a larger misorientation, than the first (local) maximum of the $f_{ori}$ versus $\xi_0 |\nabla\theta|$ relationship, may relax towards the local minimum, unless the orientational noise prevents settling into this local minimum.

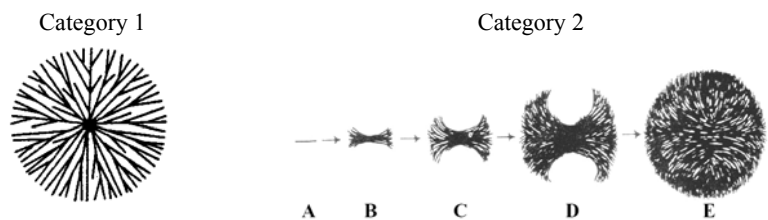

**Figure 24.** Concepts for the formation of Category 1 and 2 spherulites. From left to right: Category 1 spherulite formed via central multidirectional growth. Formation of Category 2 spherulite from a folded-chain single crystal (A) to the fully developed spherulite (E) via unidirectional growth and low angle branching [11]. Note that the latter mechanism may lead to the formation of two 'eyes' (uncrystallized holes) on the sides of the nucleation site.

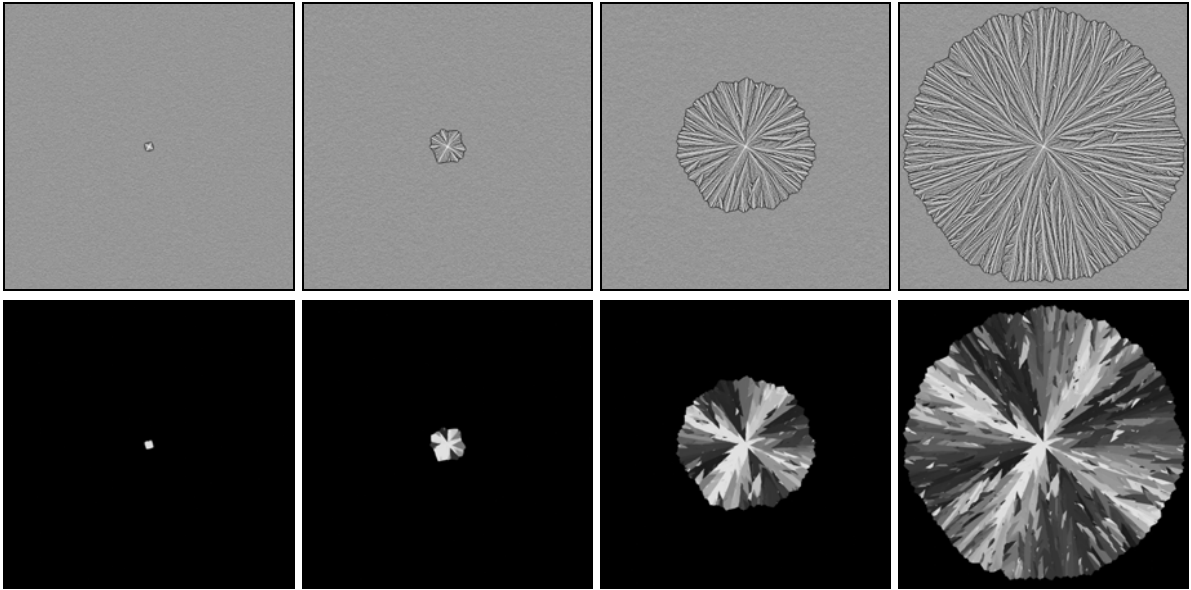

**Figure 25.** Isothermal transition between a square-shaped single crystal and a Category 1 spherulite induced by growth front nucleation, as predicted by the phase field theory (Model C). Note the gradual morphological transition, and the lack of a sharp demarcation line between areas solidified with square and spherulitic morphology in the fully-grown spherulite. With increasing size, the shape becomes more isotropic due to the randomizing effect of the newly formed grains. Note also the self-organized selection of grains whose maximum growth direction is perpendicular to the interface, yielding a cross-like pattern of grains with equivalent crystallographic orientations. ($4000 \times 4000$ grid. Snapshots taken at 500, 2000, 7500, and 15,000 dimensionless time-steps, respectively are displayed.) Upper row: composition maps (a grayscale color map was employed to increase contrast: black – liquidus, white – solidus). Lower row: orientation maps (Grayscale is the same as for the right panel of Fig. 6, except that to enhance the visibility of the pattern, the liquid regions are colored black).

Category 1 spherulites have been seen to form from transient single crystal nuclei [219]. Model C captures the gradual transition from square-shaped single crystals to circular shape under isothermal conditions. As seen in simulation, square-shaped single crystals nucleate after an initial incubation period. After exceeding a critical size (that depends on the ratio $\chi$ of the rotational and translational diffusion coefficients), the growing crystal cannot establish the same crystallographic orientation along its perimeter. Thus new grains form by growth front nucleation [196] as described in the introduction. This process gradually establishes a circular perimeter for large particles (Fig. 25).

Many studies of the early stages of spherulite growth, especially in polymers, indicate that these structures initially grow as slender thread-like fibers [8 – 11, 193, 194, 209 – 224]. These structures successively branch to form space-filling patterns. A large kinetic anisotropy ($\delta_0 = 1.99$) of two-fold symmetry is assumed, as this is expected in polymeric systems that have the propensity to form crystal filaments. Otherwise, properties of the familiar Ni-Cu system are used, as many of this systems model parameters are known, and the phase diagram is particularly simple. We include a preferred misorientation angle of 30 degrees ($m = 3$ and $x = 0.15$). The resulting growth morphologies are shown as a function of supersaturation in Fig. 26. Ideally, in a system where filament branching happens with a 30 degree misfit, the polycrystalline growth form may consist of only grains that have six well-defined orientations (including the one that nucleated), which differ by multiples of 30 degrees. This effect is especially pronounced at low supersatuations, while at high supersaturations noise-driven faults randomize the local orientation. At low supersaturations, needle crystals form. With increasing driving force, the branching frequency increases, and more space filling patterns emerge, while the average grain size decreases. This leads to a continuous morphological transition that links the needle-crystals forming at low supersaturation to axialites, to crystal sheaves, and eventually to category 2 spherulites (with 'eyes' on the two sides of the nucleus) that form at far from equilibrium. We see (second row Fig. 26) that the 'eyes' become increasingly small with increasing supersaturation, due to the increase in GFN.

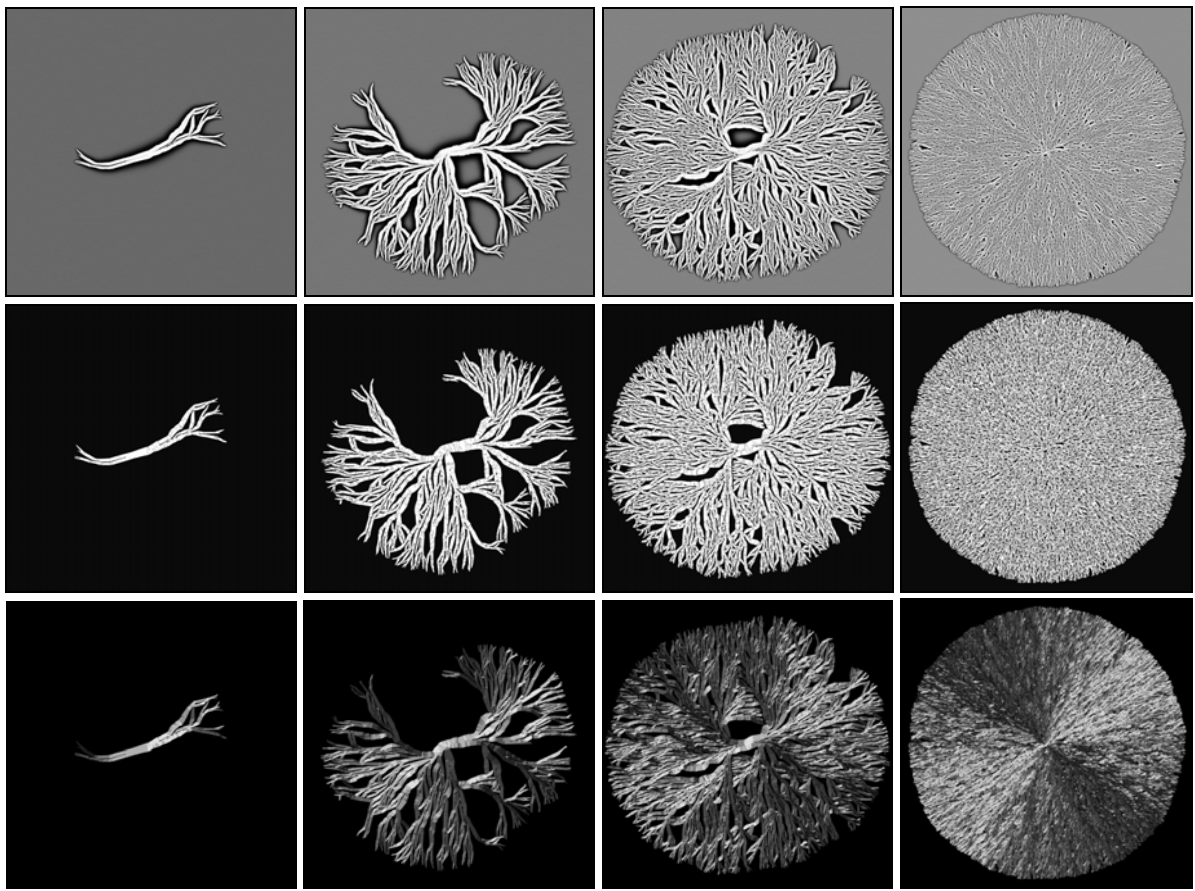

**Figure 26.** Polycrystalline morphologies formed by random branching with a crystallographic misfit of 15 degrees (Model C). The kinetic coefficient has a two-fold symmetry and a large, 99.5%, anisotropy, expected for polymeric substances. Simulations were performed on a 3000 × 3000 grid (39.6 μm × 39.6 μm). Upper row: composition map (light gray: solidus, dark gray: liquidus). Central row: grain boundary map (the gray scale in solid [crystal] shows the local orientational free energy density $f_{ori}$). Lower row: orientation map. (The shading of the orientation map is an adaptation of the scheme shown in previous figures for two-fold symmetry: When the fast growth direction is upwards, 60, or 120 degrees left, the grains are colored middle, dark or light gray, respectively, while the intermediate angles are denoted by a continuous transition among these hues. Owing to two-fold symmetry, orientations that differ by 180 degree multiples are equivalent.) Unless noise intervenes, twelve different orientations are allowed, including the orientation of the initial single crystal nucleus, which was set common for all simulations (30 degrees off horizontal direction [light gray]). In order to make the arms better discernible, in the orientation map, the liquid (which has random orientation, pixel by pixel) has been colored black. The supersaturation varies from left to right as $S$ = 0.85, 0.875, 0.90 and 0.95. Note that the branching frequency increases with increasing driving force.

Next, the time evolution of a Category 2 spherulite is considered at a fixed supersaturation (Fig. 27). First, fibrils form and then secondary fibrils nucleate at the growth front to form crystal 'sheaves'. The diverging ends of these sheaves subsequently fan out with time to form eyes, and finally a roughly spherical growth form emerges. This progression of spherulitic growth is nearly universal in polymeric materials [11, 209, 217].

What characterizes the difference between Category 1 and 2 spherulites? For category 1 spherulites, isotropy is achieved rapidly. In Fig. 25, the initial crystal had a 4-fold symmetry, and the high frequency of GFN and the associated branching leads to isotropic growth. Thus, disorder disrupts the crystalline anisotropy early in the growth process, yielding category 1 spherulites. In Fig. 27 the initial growth is fibrillar, in contrast with Fig. 25, and it takes much longer, at the same level of supersaturation (and consequent GFN), for this randomization to occur. The occurrence of Category 2 spherulites is directly related to the prevalence of early-stage fiber-type growth in comparison with the branched growth. In addition, as we increase the driving force, the time at which the growth becomes isotropic on average decreases and the structural differences between Category 1 and 2 spherulites diminish.

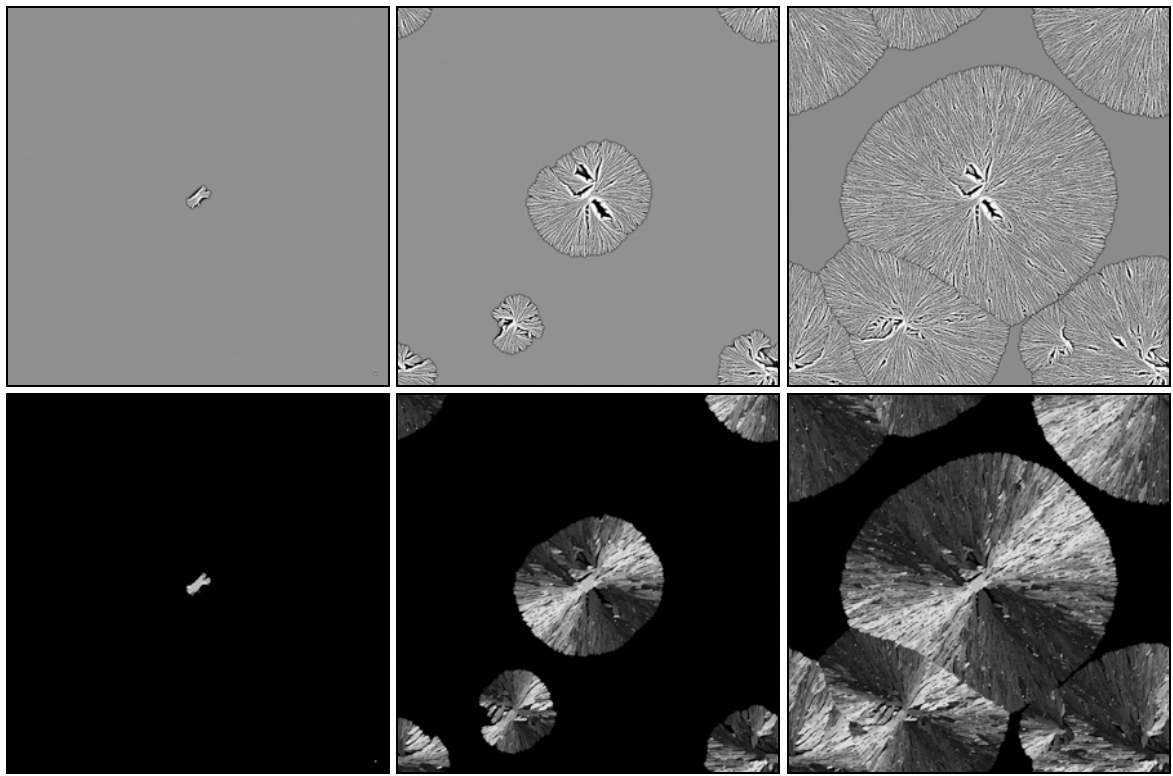

**Figure 27.** The birth of a Category 2 spherulites at unit supersaturation, in the phase field theory (Model C). Time increases from left to right. (Snapshots taken at 4.2, 8.4, 12.6, 33.5 μs after nucleation are shown. The dimensionless time used in the calculations has been transformed to real time using the diffusion coefficient of liquid Ni-Cu: $D_{NiCu} = 10^{-5}$ cm²/s. For other diffusion coefficients $D$, the times presented here have to be multiplied by $D_{NiCu}/D$.) Upper row: composition map; lower row: orientation map. Coloring and other conditions are as for fifth column in Fig. 26.

In which systems are these growth patterns prevalent? Category 1 spherulites, are a normal mode of growth in metallic and mineral systems, where fibrous growth is relatively rare. On the other hand, category 2 spherulites are ubiquitous in polymeric systems. In such fluids, high supercoolings are readily attained due to their complex molecular structure, and the fiber growth habit is characteristic of the chain-folding mechanism by which polymers crystallize [8 – 11, 209 – 224].

Category 1 and 2 spherulites may form under the same experimental conditions. How can this be understood? The early stage of growth strongly influences the late stage morphology of the spherulite. Under circumstances where the initial growth form is perturbed by fluctuations, an admixture of Category 1 and 2 spherulites is obtained. For example, simultaneous nucleation of several orientations within the same nuclei should generally yield Category 1 spherulites, but such events may be rare, and so the structures will coexist with Category 2 spherulites. Such multi-orientation nucleation events have been found in experiments on silica embedded silver particles [199] and by atomistic simulations for simple liquids [200]. Multiple nucleation events have been observed in atomic force microscopy measurements of polymer spherulite formations in thin films [220 – 222].

In the growth of compact space-filling spherulites chemical or thermal diffusion plays a negligible role. Under these conditions, the time evolution of the extent of crystallization $X$ follows the Johnson-Mehl-Avrami-Kolmogorov (JMAK) scaling $X = 1 - exp\{-[(t - t_0)/\tau]^p\}$, where $t_0$ is an incubation time due to the relaxation of the athermal fluctuation spectrum, $\tau$ is a time constant related to the nucleation and growth rates, and $p = 1 + d$ is the Avrami-Kolmogorov exponent, while $d$ is the number of dimensions [120]. Accordingly, for constant nucleation and growth rates in an infinite 2D system $p = 3$ applies. We investigated the transformation kinetics for noise-induced nucleation in a relatively large system (5,000 × 5,000 grid). To avoid the unnatural starting transient emerging from noiseless initial conditions (constant phase and concentration fields), first we heat-treated the system at 1595 K (above the liquidus curve) for 10,000 time steps, then we quenched it to 1574 K. The results are shown in Fig. 28. Fitting the JMAK relationship to the simulation data between $0.01 < X <$

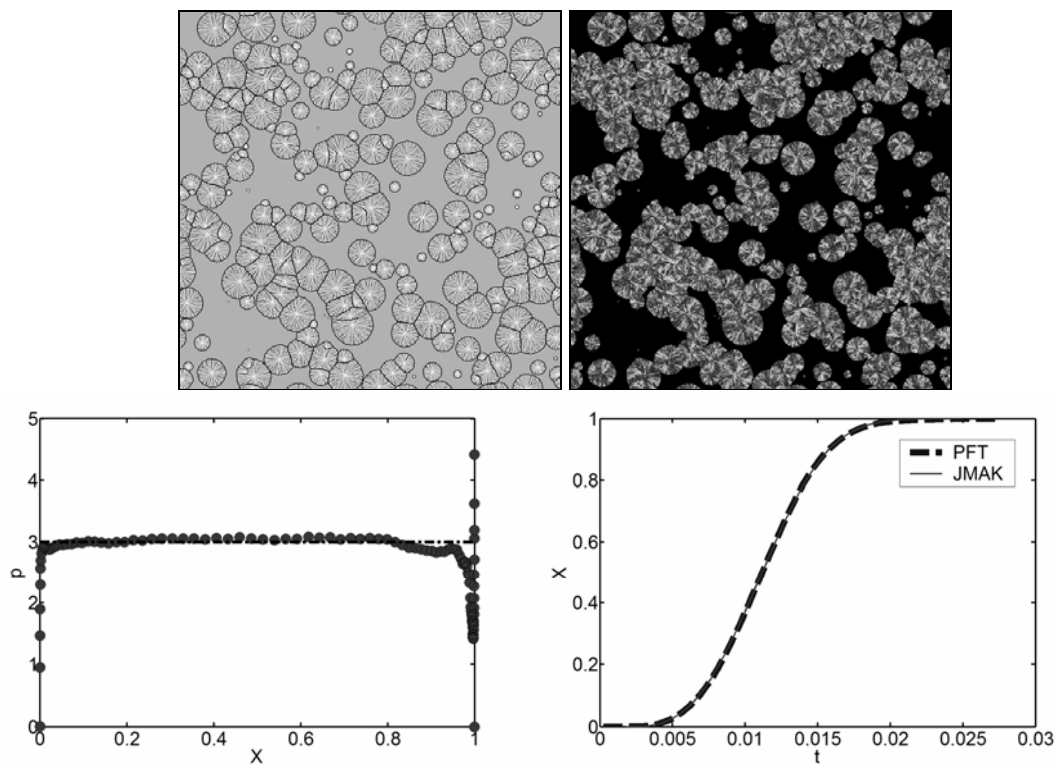

**Figure 28.** Nucleation and growth of polycrystalline spherulites in the phase field theory (Model C). (Simulation on a 5,000 × 5,000 grid). Upper row: Left concentration map; right orientation map. Lower row: Left: Transformed fraction vs. dimensionless time (dashed line), JMAK curve with the best-fit parameters (solid line). Right: Avrami-Kolmogorov exponent as a function of crystalline fraction.

0.95 (where the data are the least noisy), we find $p = 3.04 \pm 0.02$ (and $\tau = 0.0106 \pm 0.00005$, $t_0 = 0.00178 \pm 0.00005$), which is reasonably close to the $p = 3$, expected for such a transition [120].

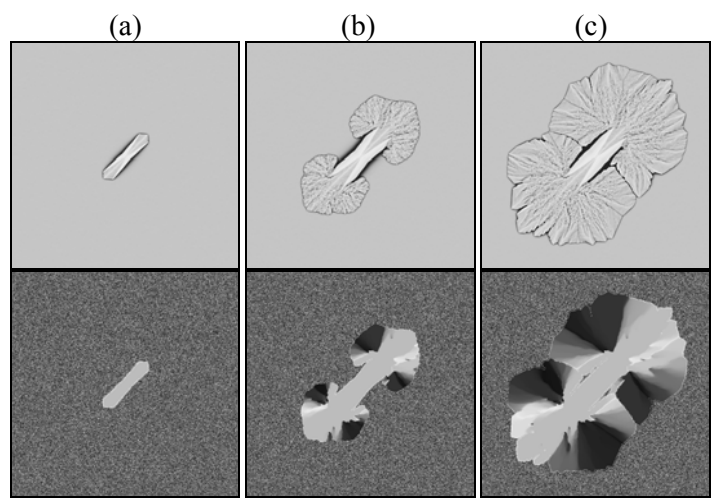

**Figure 29.** Multistage heat treatments involving spherulitic solidification, as predicted by the phase field theory (Model C): Transition between a faceted crystal habit; (a) nucleated at 1575 K] and (b) a spherulitic array after the sample is quenched to, and crystallized isothermally at 1571 K ($M_\theta$ is reduced by a factor of 20), and back to faceted growth (c) after returning to 1575 K. (Compared to polymers, this system requires a relatively small temperature cycling range due to the ideal solution behavior of the Ni-Cu system.) Note the formation of new crystal grains due to GFN during the low temperature stage of the cycling. The computations were performed with 5% anisotropy of the interface free energy (of six-fold symmetry) and 85% anisotropy of the phase field mobility (of two-fold symmetry) on a 1000 × 1000 grid. Upper row: composition maps (black: liquidus, light gray: solidus). Lower row: orientation maps (grayscale as in Fig. 6).

(a)  (b)

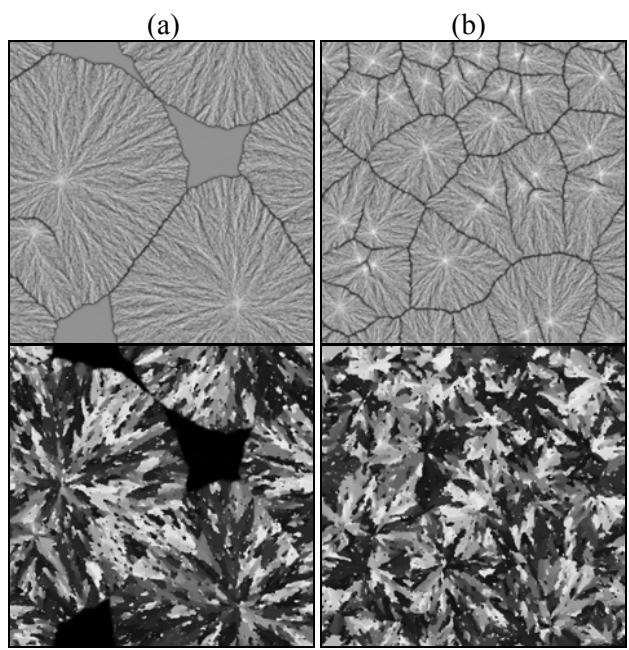

**Figure 30.** Multistage heat treatments involving spherulitic solidification, as predicted by the phase field theory (Model C): Morphologies formed after two thermal routes are shown which have the same final temperature. In (a) spherulitic solidification occurs at 1574 K after direct quenching from above the melting point (1595 K). In (b) spherulitic solidification occurs at 1574 K, after deep quenching first to 1350 K. Note the similarity of the growth forms, and the enhanced number of crystallites in the latter case. The computations were performed with 10% anisotropy of the interface free energy (of four-fold symmetry) on a 500 × 500 grid. Upper row: composition maps (black: liquidus, light gray: solidus). Lower row: orientation maps (grayscale as in Fig. 6).

(a)  (b)  (c)

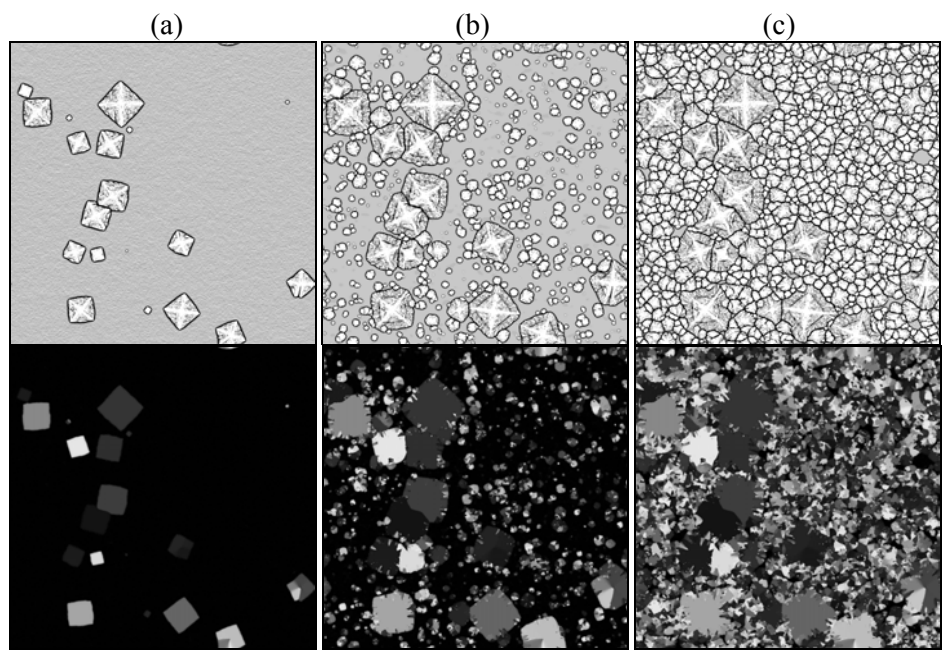

**Figure 31.** Multistage heat treatments involving spherulitic solidification, as predicted by the phase field theory (Model C): In (f)–(h) spherulitic overgrowth occurs on pre-existing square crystals with parallel nucleation and growth of spherulites. Square-crystals were formed at 1574 K isothermally, then quenched to 1570 K where crystallization completed. The computations were performed with 15% anisotropy of the interface free energy (of four-fold symmetry) on a 1000 × 1000 grid. Upper row: composition maps (coloring: a grayscale colormap was employed to increase contrast: black – liquidus, white – solidus). Lower row: orientation maps (grayscale as in Fig. 6).

There is a great deal of interest in how temporal variations in processing conditions (temperature, pressure, etc.) influence spherulitic growth morphology. Multistage heat treatments on polymeric substances have demonstrated that that both the local growth morphology and growth rate depend on the temperature, but are independent of previous thermal history [194, 212, 219]. For example, cycling between two temperatures reversibly switches between faceted and spherulitic growth morphologies both in experiment [194, 219] and simulation [Fig. 29(a)-(c)]. The predominance of either growth morphology depends on the cycling time, and complex patterns are generated in this fashion. For example, following experiment [219], we can simulate either a direct quench to the temperature of spherulitic solidification from above the melting point or instead simulate a deeper quench followed by heating to the same final temperature. As shown in Figs. 30(a),(b) these different histories yield much the same late stage growth form, but a larger number of spherulites in the latter deep quench case (due to enhanced nucleation at lower temperatures). Finally, other experiments [219] show that spherulitic overgrowth occurs on square-shaped crystals grown at small undercoolings, while, simultaneously, normal spherulites fill the remaining space. This behavior is recovered by our phase field simulations [Fig. 31(a) − (c)]. The ability of Model C to reproduce such complex sequences suggests that our field theory contains the essential physics necessary to describe a broad range of real materials.

We now return to the wide range of spherulitic crystallization patterns shown in Fig. 1. Can Model C explain this variability? Fig. 32 shows a selection of simulations that bear resemblance to the experimentally observed morphologies [7 − 9, 11, 193, 211, 223 − 225]. In addition to the category 1 and 2 spherulites described above, we observe structures ranging from spiky and arboresque spherulites, to 'quadrites' [11, 223] exhibiting a cross-hatching fine structure [see Fig. 1(d)], to undulating branched patterns. These simulations differ only in the driving force, anisotropies, branching angle, and mobilities, indicating that the essential features of a broad variety of spherulitic morphologies can be captured, using only a few model parameters. Another interesting polycrystalline morphology is the disordered fractal-like growth form seen in electrodeposition [12] [Fig. 1(i)]. With appropriate choice of model parameters this pattern can also be recovered (Fig. 33).

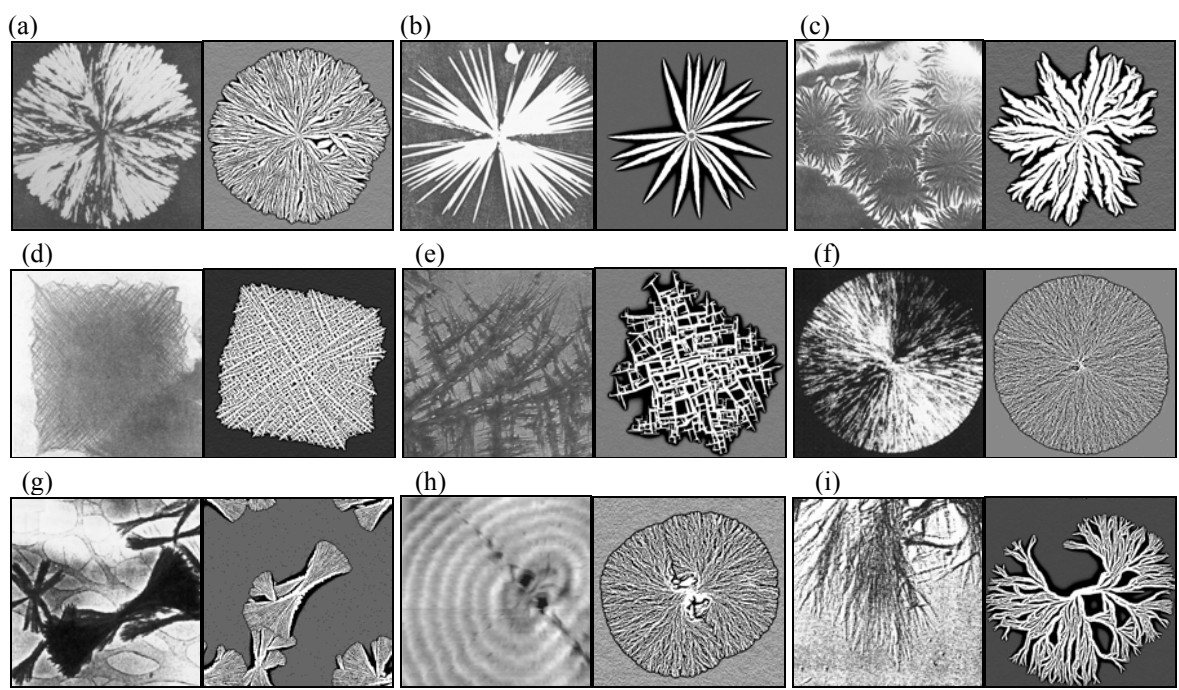

**Figure 32.** A selection of polycrystalline growth morphologies vs. patterns from phase field simulations (Model C): The experimental and theoretical images are arranged into pairs. Left: experiment; right: simulation. Note that with a few macroscopic parameters (anisotropy of the interface free energy and kinetic coefficient, branching angle, and depth of the metastable well of the orientational free energy) a broad variety of solidification patterns can be captured. (The experimental images originate from the following works: (a) reprinted from [224], © 1964, with the permission of the AIP; (b) reprinted from [193], © 1963, with the permission of the AIP; (c) reprinted from [225]; (d) reprinted from [11]; (e) reprinted from [223], © 1986, with the permission of Wiley; (f) reprinted from [7], © 1988, with the permission of Elsevier; (g) reprinted from [8], © 1993, with the permission of ACS; (h) reprinted from [211], © 2000, with the permission of Elsevier; (i) reprinted from [9], © 1965, with the permission of AIP.)

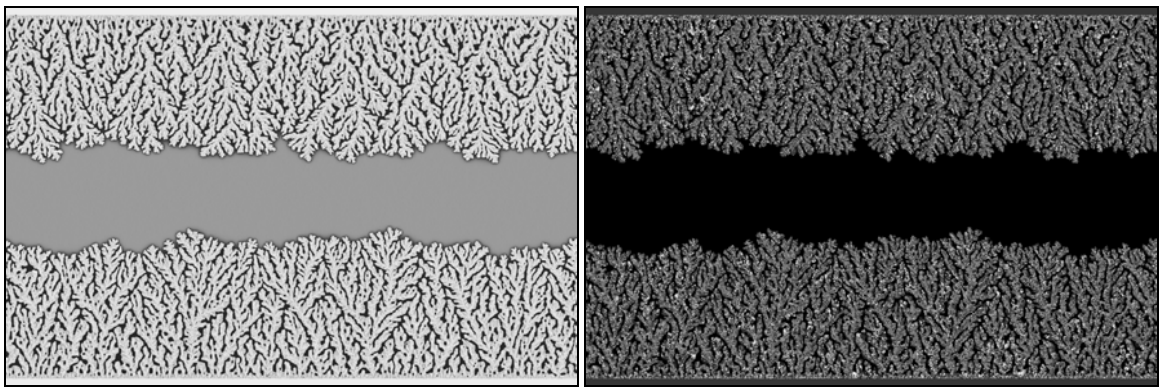

**Figure 33.** Fractal-like polycrystalline aggregate grown on planar single crystal surfaces as predicted by the phase field theory (Model C). The branching angle is 60°, $x = 0.3$, supersaturation of 0.9, while a 99.5% anisotropy and a twofold symmetry of the kinetic coefficient are assumed. The calculation has been performed on a rectangular grid of size 3000 × 2000. The composition and orientation maps are shown (up and down, respectively). Grayscale is the same as for Fig. 26.

Work is underway to map the zoo of possible polycrystalline morphologies. While the similarity of the simulations and the experimental patterns is reassuring, further experimental work is also needed to determine, whether the predicted grain structures are indeed realistic. The study on spherulites summarized above has been done by the present authors in collaboration with J. A. Warren and J. F. Douglas from the National Institute of Standards and Technology, Gaithersburg. Details will be presented elsewhere [229].

## 2.7. Crystallization in the Presence of Walls

Solidification in the presence of walls is of great practical importance. In casting, solidification usually starts by heterogeneous crystal nucleation on the walls of the mould (see e.g., [230]). With the exception of extremely pure samples, even volume nucleation happens mostly via a heterogeneous mechanism (on the surface of floating foreign particles) [231]. Particulate additives are widely used as grain refiners, to reduce grain size by enhancing the nucleation rate. Nonetheless, heterogeneous nucleation is probably the only stage of solidification where the micro-mechanism of the process is largely unknown. (An exception is the inoculation of metallic alloys, where the number of grains is controlled by the free growth condition rather than heterogeneous nucleation itself [232 − 234].) While condensation on solid substrate or in slit pore has been addressed with advanced methods [235 − 238], fewer investigations have been performed for heterogeneous crystal nucleation.

Although the phase field method has been used to address problems that incorporate heterogeneous nucleation, this process is usually mimicked introducing supercritical particles into the simulation window [70, 71]. Recently, however, steps have been made towards a physical modeling of heterogeneous nucleation within the phase field theory. Castro [174] introduced walls into a single order parameter theory by assuming a no-flux boundary condition at the interface ($\mathbf{n}\nabla\phi = 0$, where $\mathbf{n}$ is the normal vector of the wall), which results in a contact angle of 90 degrees at the wall-solid-liquid triple junction. Langevin noise is then introduced to model nucleation. It has been found that the presence of walls enhances nucleation, and thus the internal corners are places where nucleation is more likely to occur.

Prescribing ($\mathbf{n}\nabla\phi$) = 0 and ($\mathbf{n}\nabla c$) = 0 at the wall perimeter, Gránásy *et al.* introduced chemically inert surfaces into Model A, and performed simulations to address heterogeneous volume nucleation on foreign particles (a more detailed description than the 'pinning centers'), on rough surfaces, and in confined space (porous matter and channels) [33, 238]. A few preliminary results, which illustrate that various complex phenomena can be addressed this way, are shown in Figs. 34 − 40. These include noise-induced nucleation of dendritic crystals on square-shaped particles and rough surfaces, particle engulfment, solidification in porous medium, and in a rectangular channel (orientation selector), in two and three dimensions. In the 3D calculations, a simpler model is used, that does not handle the differences in the crystallographic orientation (Model A without orientation field).

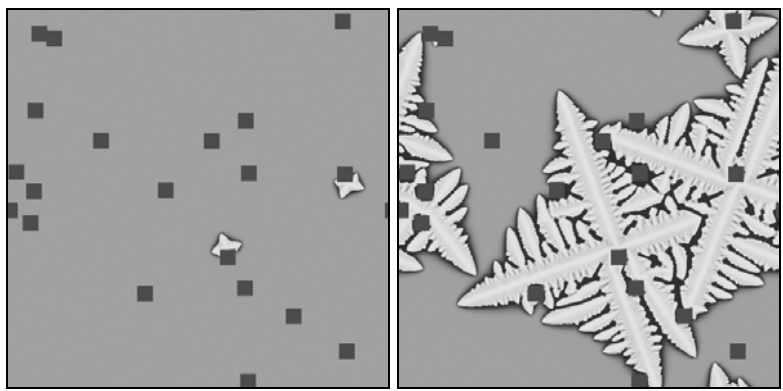

**Figure 34.** Heterogeneous nucleation of dendritic crystals on square-shaped foreign particles as predicted by the phase field theory (Model A). Composition maps are shown. Coloring: dark gray: foreign particles; black: liquidus; white: solidus; light gray: initial liquid.

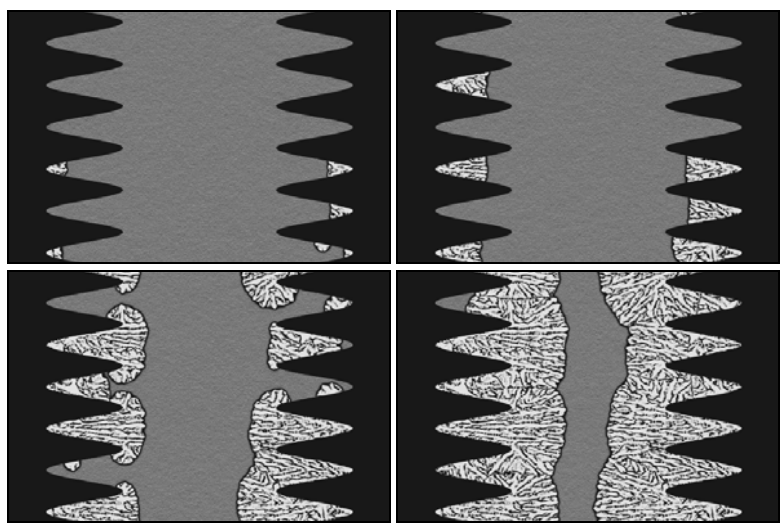

**Figure 35.** Heterogeneous nucleation on rough surfaces as predicted by the phase field theory (Model A). Composition maps are shown. Black: substrate; white: solid; grays: liquid. To enhance the visibility of details, the contrast of the images has been increased.

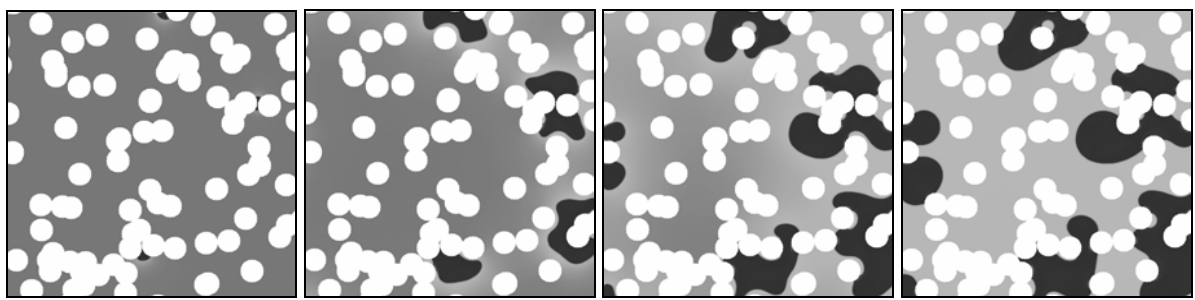

**Figure 36.** Heterogeneous nucleation and growth of stoichiometric crystals in porous matter as predicted by Model A. Chemical composition maps are shown (white: particles of porous matter, black: crystal, gray: liquid). (The model describes the formation of solid $CO_2$ hydrate in a supersaturated aqueous $CO_2$ solution at $T = 274$ K, and at a pressure of $P = 15$ MPa.) Note that nucleation happens in the notches between particles of the porous matter, and the depletion zone forming around the growing crystals.

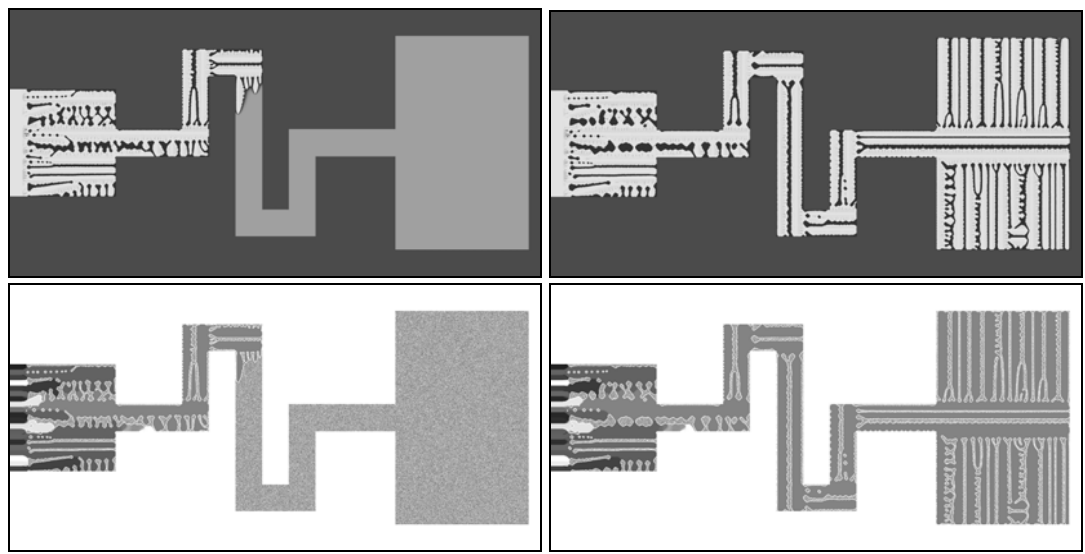

**Figure 37.** Dendritic solidification in a two-dimensional orientation selector (pigtail) mimicking the casting of single crystal components as predicted by Model A. Top: composition field (dark gray: mould; black: liquidus; white: solidus; light gray: initial liquid); bottom: orientation field (grayscale is the same as in right panel of Fig. 6, white: mould).] Crystallization starts on the left with several crystallographic orientations, but only a single crystallographic orientation survives the meandering channel to reach the liquid volume on the right in the simulation box.

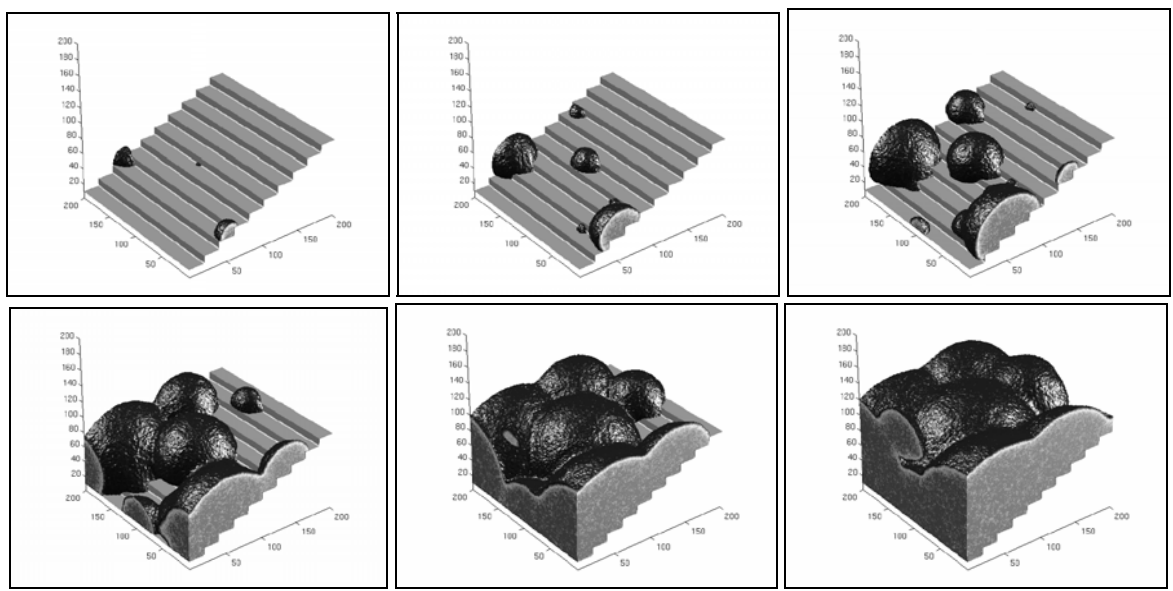

**Figure 38.** Phase field calculations for noise-induced heterogeneous crystal nucleation on stairs in three dimensions. Properties of Ni and the physical interface thickness (1 nm) are used. The calculation was performed at 1200 K on a 200×200×200 grid (40 nm×40 nm×40 nm). Note that as expected nucleation happens at the inside corners of the stairs, a position energetically preferred. The shiny black surface corresponds to the solidification front ($\phi = \frac{1}{2}$), while the light spotted gray regions denotes bulk crystalline properties.

Heterogeneous noise-induced nucleation has been investigated in various geometries including a stair-like surface, porous medium (represented by cubes placed on a bcc lattice), and 3D checkerboard-like modulated surface (Figs. 38 – 40). Such studies may contribute to a better understanding of processes that can be used in micro/nano-patterning. Future work will explore the kinetics of such processes, and extend the modeling to arbitrary contact angles.

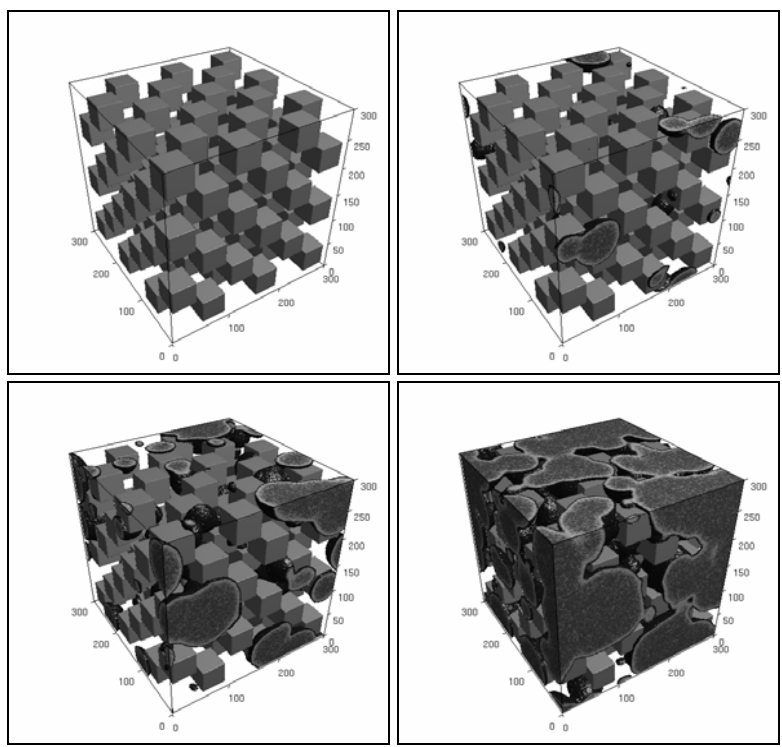

**Figure 39.** Phase field calculations for noise induced heterogeneous crystal nucleation on cubic particles arranged on a bcc lattice. Properties of Ni and the physical interface thickness (1 nm) are used. The calculation was performed at 1200 K on a 300×300×300 grid (60 nm×60 nm×60 nm).

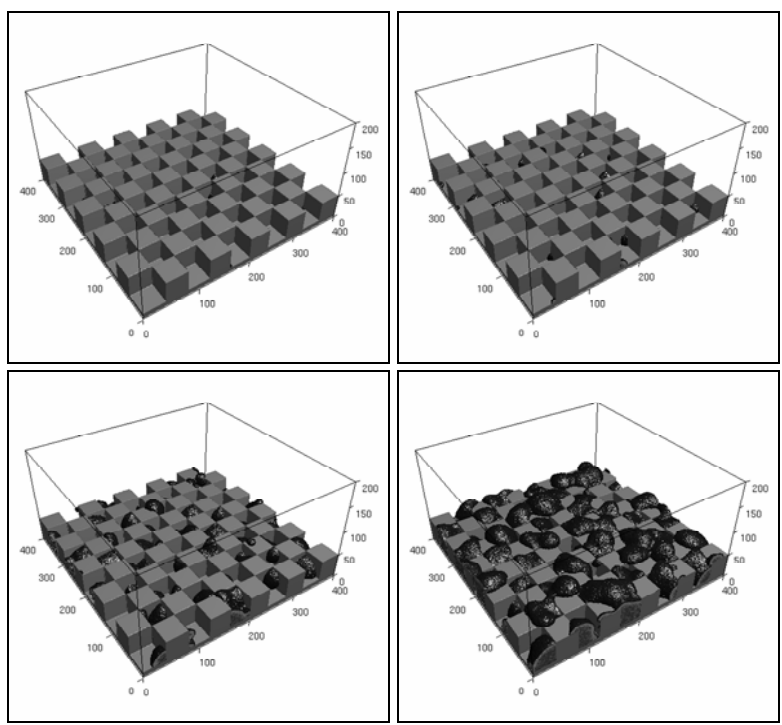

**Figure 40.** Phase field calculations for noise induced heterogeneous crystal nucleation on a 3D checkerboard-modulated surface. Properties of Ni and the physical interface thickness (1 nm) are used. The calculation was performed at 1200 K on a 400×400×200 grid (80 nm×80 nm×40 nm).

# 3. SUMMARY AND FUTURE DIRECTIONS

In the previous section we demonstrated the capability of the phase field method to describe complex polycrystalline morphologies. This includes nucleation and growth problems in anisotropic systems, the effect of particulate additives and trapped disorder on growth morphologies, solidification in confined space, and many others. However, systematic studies are needed on all areas discussed here (e.g., mapping of possible morphologies, study of transformation kinetics in the presence of walls, etc.). Of particular interest to us is the validation of the models that include random branching. This requires the collection of statistics on the morphologies both in theory and experiment, as these patterns may have only statistical similarity. Straightforward extensions of the work performed with Model A to C may include coupling to hydrodynamics (an essential step to study particle-front interaction) as done for simpler cases [21, 106, 239 − 246], and dynamically changing temperatures [247, 248]. Interesting further directions for polycrystals are the modeling of crack dynamics [249, 250], and the addressing of reaction-diffusion type phenomena such as oxidation, hydrogenation, electrodeposition [251 − 253], etc.

The largest theoretical challenge is perhaps the generalization of the model to three dimensions. This requires minimum three orientational fields (e.g., two polar angles that set the fast growth direction in 3D, and a third angle that specifies the rotation of the crystal around this axis). Recent work offers two equivalent solutions to this problem [254 - 257]. Kobayashi and Warren [254, 256] addressed grain boundary dynamics in 3D, while Pusztai *et al.* [255,257] describe polycrystalline solidification in 3D, including the growth of several dendritic particles (Fig. 41) and the nucleation and growth of needle crystals with different crystallographic orientations and the formation of polycrystalline spherulites, sheaves (Fig. 41) and seaweed-like patterns.

The simplistic model of elasticity that these models inherently contain should be refined using the continuously developing inventory of phase field models for solid-state transformations [93, 258–264]. The broad interface remains an issue (enhanced solute trapping, etc.). New approaches (e.g., [265]) should address some of the numerical issues associated with too thick interfaces, while interaction with atomistic scale modeling will help to fix the model parameters for quantitative calculations.

Simulations with "no-flux" walls represent only a significantly simplified description of the interaction of crystallization with foreign solid surfaces (substrate). A full treatment of heterogeneous nucleation has to incorporate the chemical interactions, shall include an extra phase field for the substrate, while the coefficients of the associated gradient terms will incorporate the energetics that determines the contact angle at the liquid-crystal-substrate junction.

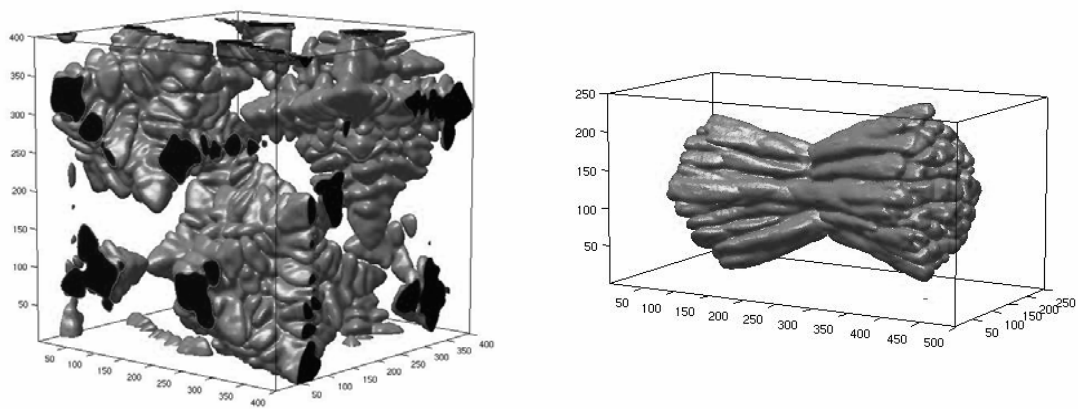

**Figure 41.** Phase field simulations in 3D for the growth of four randomly oriented crystal nuclei (left) with cubic crystal symmetries and for the formation of a polycrystalline sheaf (right) (by Pusztai, Bortel and Gránásy). The $\phi = 0.5$ surfaces are shown. The dendrites were grown on a 400×400×400 grid (5.25 μm×5.25 μm×5.25 μm) with cubic crystal symmetries and a corresponding kinetic anisotropy, while the simulation for the sheaf has been performed on a 500×250×250 grid (6.56 μm×3.28 μm×3.28 μm) with triclinic crystal symmetries and an ellipsoidal anisotropy of the kinetic coefficient. Note the effect of periodic boundary conditions: the dendrite arms growing out of the simulation box on one side enter on the opposite side.

Regarding future directions, we mention a most remarkable new approach to polycrystalline solidifications by Elder *et al.* [266] that allows for atomistic modeling of the solid-liquid transition on diffusive time scales. A negative square gradient term is balanced here by a positive fourth order term, yielding homogeneous (liquid) regions and crystalline ('dots' on lattice) regions in the phase diagram, with a first order phase transition in between. The model naturally incorporates crystal anisotropy, elastic and plastic deformations, grain boundaries, cracks, epitaxy, etc. [266, 267]. Its results are consistent with the Read-Shockley theory of grain boundary energy and the Matthews-Blakeslee theory for misfit dislocations in epitaxy [266, 267]. The model has been applied for eutectic solidification and dendritic growth [268]. A computationally efficient coarse-grained field theory, equivalent with this atomistic phase field model, has been developed using renormalization technique by Goldenfeld *et al.* [269]. With the ever-increasing power of computers, these approaches are expected to take over many of the tasks of the conventional phase field theories, provided that their 3D generalizations turns out to be as successful as the 2D forms.

Finally, one should mention recent advances made using the multi-phase-field approach in modeling various aspects of polycrystalline solidification [270-273]. Further work is, however, needed (i) to clarify third phase appearance at the phase boundaries, and (ii) to consider crystal symmetries in grain boundary formation/evolution (a problem successfully addressed by phase field approaches with orientation fields in 2D and 3D [141-144,255,257]).

Summarizing, we believe that, in the foreseeable future, phase field theories, and field theoretic methods in general, will remain among the most powerful tools of computational materials science. They will significantly contribute to our qualitative understanding of crystallization processes including nucleation, and will become useful in quantitative modeling of the formation and evolution of crystalline microstructure from the nanometer scale to the micron or even to the millimeter scale.

# ACKNOWLEDGMENTS

L. G. is indebted to J. A. Warren and J. F. Douglas (National Institute of Standards, Gaithersburg, Maryland) for the collaboration and the illuminating discussions on phase field theory and polycrystalline solidification. The authors thank K. Lee, W. Losert, B. D. Nobel, P. F. James, V. Ferreiro, J. F. Douglas, G. Ryshchenkow, G. Faivre, M. Ojeda, D. C. Martin, F. Khoury, B. Lotz, and V. Fleury for the experimental images. L. G. is grateful to M. Plapp for the extensive discussions on the phase field theory. This work has been supported by contracts OTKA-T-037323, ESA PECS Contract No. 98005, and by the EU FP6 Integrated Project IMPRESS under Contract No. NMP3-CT-2004-500635 (funding institutions: the European Commission and the European Space Agency), and forms part of the ESA MAP Projects No. AO-99-101 and AO-99-114. T. P. acknowledges support by the Bolyai János Scholarship of the Hungarian Academy of Sciences.

# APPENDIX

We specify here three models (Models A, B and C), which show rather similar features, however, differ in important details such as the form of the free energy functional. *Note that the models termed here as Models A to C differ from Models A to C of the usual Hohenberg-Halperin [37] classification.*

## A.1 Model A

The local state of matter is characterized by the phase field $\phi$. This order parameter describes the extent of structural change during freezing and melting. The other basic field variables are the chemical composition $c$ and the normalized orientation field $\theta$ [99, 100], where $\theta$ specifies the orientation of crystal planes in the laboratory frame. The free energy $F$ consists of various contributions that will be discussed below:

$$F = \int d^3r \left\{ \begin{array}{l} \dfrac{\varepsilon_\phi^{\,2} T}{2} s^2(\vartheta - \theta)\left|\nabla\phi\right|^2 + \dfrac{\varepsilon_c^{\,2} T}{2}\left|\nabla c\right|^2 + \\[2mm] + w(c)T g(\phi) + [1-p(\phi)][f_S(c,T) + f_{ori}(\left|\nabla\theta\right|)] + p(\phi) f_L(c,T) \end{array} \right\}, \qquad (A1)$$

where

$$\varepsilon_\phi^{\,2} = \frac{6\sqrt{2}\gamma_{A,B}\delta_{A,B}}{T_{A,B}}, \quad w(c) = (1-c)w_A + c w_B, \quad w_{A,B} = \frac{12\gamma_{A,B}}{\sqrt{2}\delta_{A,B}T_{A,B}},$$

$$g(\phi) = \tfrac{1}{4}\phi^2(1-\phi)^2, \quad g'(\phi) = \phi^3 - \tfrac{3}{2}\phi^2 + \tfrac{1}{2}\phi$$

$$p(\phi) = \phi^3(10 - 15\phi + 6\phi^2), \quad p'(\phi) = 30\phi^2(1-\phi)^2,$$

$$f_{ori} = HT\left|\nabla\theta\right|$$

$$s(\vartheta - \theta) = 1 + s_0\cos(k\vartheta - 2\pi\theta), \quad \vartheta = \arctan\left[(\nabla\phi)_y / (\nabla\phi)_x\right]$$

Here $\varepsilon_\phi$ and $\varepsilon_c$ are the coefficients of the square-gradient term for the fields $\phi$ and $c$; $w_i$ is the free energy scale for the i-th pure component (i = A, B); $s$, $g$ and $p$ are the anisotropy function, the quartic double-well function and the interpolation function. $\gamma_i$, $\delta_i$, $T_i$ are the interface free energy, interface thickness and melting point for the i-th pure component (i = A, B). $\vartheta$ is the inclination of the normal vector of the interface in the laboratory frame. $H$ determines the free energy of the low angle grain boundaries. $s_0$ is the amplitude of the anisotropy of the interface free energy, while $m$ is the symmetry index ($m = 6$ for six-fold symmetry). Time evolution is governed by relaxational dynamics and Langevin noise terms are added to model thermal fluctuations [99, 100],

$$\dot{\phi} = -M_\phi \frac{\delta F}{\delta\phi} + \zeta_\phi = M_\phi \left\{ \nabla\left(\frac{\partial f}{\partial\nabla\phi}\right) - \frac{\partial f}{\partial\phi} \right\} + \zeta_\phi$$

$$\dot{c} = \nabla M_c \nabla\left(\frac{\delta F}{\delta c} - \zeta_j\right) = \nabla\left\{ Dc(1-c)\nabla\left[\left(\frac{\partial f}{\partial c}\right) - \nabla\left(\frac{\partial f}{\partial\nabla c}\right) - \zeta_j\right] \right\}, \qquad (A2)$$

$$\dot{\theta} = -M_\theta \frac{\delta F}{\delta\theta} + \zeta_\theta = M_\theta \left\{ \nabla\left(\frac{\partial f}{\partial\nabla\theta}\right) - \frac{\partial f}{\partial\theta} \right\} + \zeta_\theta$$

where $\zeta_i$ are the appropriate Langevin-noise terms.

The time scales for the three fields are determined by the appropriate mobilities appearing in the equations of motion, and $M_\phi$, $M_c$ and $M_\theta$ are the mobilities associated with coarse-grained equation of motion, which in turn are related to their microscopic counterparts. The mobility $M_c$ is directly proportional to the classic interdiffusion coefficient for a binary mixture. The mobility $M_\phi$ dictates the rate of crystallization, while $M_\theta$ controls the rate at which regions reorient.

### A.1.1. Phase Field

Using the length and time scales $\xi$ and $\xi^2/D_l$, respectively ($D_l$ is the chemical diffusion coefficient in the liquid), the dimensionless phase field mobility $m_\phi = m_{\phi 0}\{1 + \delta_0\cos[k(\psi - \theta)]\}$, and $m_{\phi 0} = M_\phi\varepsilon_\phi^2 T/D_l$, the following *dimensionless form* emerges

$$\widetilde{\dot{\phi}} = m_\phi \left[ \begin{array}{l} \widetilde{\nabla}(s^2\widetilde{\nabla}\phi) - \dfrac{\partial}{\partial\widetilde{x}}\left\{ s\dfrac{\partial s}{\partial\vartheta}\dfrac{\partial\phi}{\partial\widetilde{y}} \right\} + \dfrac{\partial}{\partial\widetilde{y}}\left\{ s\dfrac{\partial s}{\partial\vartheta}\dfrac{\partial\phi}{\partial\widetilde{x}} \right\} \\[3mm] - \xi^2\,\dfrac{w(c)T g'(\phi) + p'(\phi)\{f_L(c,T) - f_S(c,T) - f_{ori}\}}{\varepsilon_\phi^{\,2} T} \end{array} \right]. \qquad (A3)$$

Henceforth quantities with tilde are dimensionless, while prime denotes differentiation with respect to the argument of the function.

### A.1.2. Concentration Field

Following previous work [50, 51], we choose the mobility of the concentration field as $M_c = (v_m/RT) D c (1 − c)$, where $v_m$ is the average molar volume, and $D = D_s + (D_l − D_s) p(\phi)$ is the diffusion coefficient. This choice ensures diffusive equation of motion. Since $HT$ is assumed independent of concentration, no coupling to the orientation field emerges. Introducing the reduced diffusion coefficient $\lambda = D/D_l$, the *dimensionless equation of motion* for the concentration field reads as

$$\widetilde{\dot{c}} = \widetilde{\nabla}\left\{\frac{v_m}{RT}\lambda c(1-c)\widetilde{\nabla}\left[(w_B - w_A)Tg(\phi) + [1 - p(\phi)]\frac{\partial f_S}{\partial c}(c,T) + p(\phi)\frac{\partial f_L}{\partial c}(c,T) - \frac{\varepsilon_c^2 T}{\xi^2}\widetilde{\nabla}^2 c\right]\right\}. \quad (A4)$$

### A.1.3 Orientation Field

Introducing the dimensionless correlation length of the orientation field $\widetilde{\xi}_0 = \xi_0 / \xi$ , and defining the dimensionless orientational mobility as $m_\theta = [M_{\theta S} + (M_{\theta L} − M_{\theta S})\cdot p(\phi)]\ \xi HT/D_l$, the *dimensionless equation of motion* is as follows as

$$\widetilde{\dot{\theta}} = m_\theta\left[\widetilde{\nabla}\left\{[1 - p(\phi)]\frac{\widetilde{\nabla}\theta}{|\widetilde{\nabla}\theta|}\right\} - \frac{\varepsilon_\phi^2}{H\xi}s\frac{\partial s}{\partial\theta}\left|\widetilde{\nabla}\phi\right|^2\right], \quad (A5)$$

This form of $f_{\text{ori}}$, and the noise added to the equation of motion ensure that the orientation field $\theta$ is random in space and time in the liquid. This makes possible to quench orientational defects into the solid, leading to polycrystalline growth. Independently, branching with fixed relative misorientation may occur, i.e., sharp (step-like) grain boundaries of fixed orientational misfit (of fixed grain boundary energy) appear.

The second term on the RHS of (A5) must be handled with care. It is negligible if the physical interface thickness (~ 1 nm) is used. Due to limitations of computer power, we employ a relatively broad interface compared to those found in metallic alloys. This broad interface leads to artifacts that are not present with thinner interfaces. As a practical matter, we adopt one of the following measures: (a) perform the calculations in the presence of only kinetic anisotropy (then this term is zero); (b) we omit this term.

In *modeling eutectic and peritectic solidification* with locked orientational relationship between the two solid phases ($\alpha$ and $\beta$), we make the following additional assumptions: In solids of composition close to the equilibrium ones, the grain boundary model used for the single solid case is retained (i.e., $f_{\text{ori}} \propto |\nabla\theta|$). At the interface between the two solid phases a non-zero orientation change is preferred. This orientation change is assumed to be independent of the inclination of the solid-solid interface. To implement these, the orientation free energy is assumed to be composition dependent: $f_{\text{ori}} = [1 − p(\phi)]\ HT\cdot F(c,|\nabla\theta|, |\nabla\theta|^2)$, where $F = h(c)\ F_1(|\nabla\theta|) + [1 − h(c)]\ F_2(|\nabla\theta|) + \frac{1}{2}\ \varepsilon_c^2 T\ (HT)^{-1}|\nabla\theta|^2$. The square-gradient term for the orientation field is needed to allow for grain boundary motion (besides grain rotation [141]). We use the following set of functions

$$h(c) = \tfrac{1}{2}\ \{1 + cos[(c−c_\alpha)/(c_\beta−c_\alpha)2\pi]\},$$

$$F_1(|\nabla\theta|) = |\nabla\theta|,$$

$$F_2(|\nabla\theta|) = a + b\ |cos(2n\pi d|\nabla\theta| + \psi)|.$$

Here $c_\alpha$ and $c_\beta$ are the equilibrium solid compositions, $a$ and $b$ are constants, $d$ is the interface thickness, while $n$, or $\psi$, or both can be used to define the preferred orientation change at the interface.

Note that function $h(c)$ switches on the preference for the orientation change for *solid* compositions that differ significantly from the equilibrium solid compositions $c_\alpha < c < c_\beta$. Such intermediate compositions normally occur only at the $\alpha - \beta$ interface in the solid. Anisotropy of the interface free energy might be incorporated via making either $\varepsilon_\theta{}^2$ or $\psi$ dependent on the interface inclination angle $\vartheta_c$ $= atan[(\partial c/\partial y)/(\partial c/\partial x)]$. While the functions used here have been chosen intuitively, for different crystal structures, $F_1$ and $F_2$ can be deduced on physical grounds.

Having defined these relationships, new terms emerge in the governing equations. The actual $F$ $= h(c) F_1 + [1 - h(c)] F_2 + {}^1/{}_2 \, \varepsilon_\theta{}^2 T (HT)^{-1} |\nabla \theta|^2$ function enters into equation (A3), its first derivative $F' = h(c) F_1' + [1 - h(c)] F_2' + \varepsilon_\theta{}^2 T (HT)^{-1} |\nabla \theta|$ appears in equation (A5), and the terms shown below are added to the RHS of equation (A4).

$$\widetilde{\nabla} \lambda c(1-c) \frac{\nu}{RT} M \left\{ \begin{array}{l} - p'(\phi) \, h'(c) \big[ F_1 - F_2 \big] \widetilde{\nabla} \phi + [1 - p(\phi)] h''(c) \big[ F_1 - F_2 \big] \widetilde{\nabla} c + \\[2mm] + [1 - p(\phi)] h'(c) \Big[ F_1{}' - F_2{}' \Big] \dfrac{1}{\xi} \widetilde{\nabla} \big| \widetilde{\nabla} \theta \big| \end{array} \right\}.$$

In all cases the governing equations have been solved numerically using an explicit finite difference scheme. Unless stated otherwise periodic boundary conditions were used. Convergence of the solution with the orientational equation needs that the time step for the orientation equation be less than about 1/30 of the time step required for the stable solution of the other fields.

A parallel code has been developed that relies on the Message Passing Interface (MPI) protocol and was run on two PC clusters built up in the Research Institute for Solid State Physics which is dedicated exclusively to phase field simulations. At present, the clusters consist of 60 and 100 nodes, respectively and a server machine for each. All simulations shown in this paper have been performed on these clusters.

## A.2. Model B

Model B is a four-field extension of Model A. The extra phase field monitors the transition between solid phases (structures) $\alpha$ and $\beta$. Regular thermodynamics is used and a square-gradient term is incorporated for the concentration. The free energy functional reads as

$$F = \int d^3 r \left\{ \begin{array}{l} \dfrac{\varepsilon_\phi{}^2 T}{2} |\nabla \phi|^2 + \dfrac{\varepsilon_c{}^2 T}{2} |\nabla c|^2 + w_\phi(c) T g(\phi) \\[2mm] + [1 - p(\phi)][f_S(\psi, \nabla \psi, c, T) + f_{ori}(\psi, |\nabla \theta|)] + p(\phi) f_L(c, T) \end{array} \right\}, \qquad (A6)$$

where

$$f_S(\psi, \nabla \psi, c, T) = \frac{\varepsilon_\psi{}^2 T}{2} \big| \nabla \psi \big|^2 + w_\psi T g(\psi) + [1 - p(\psi)] f_\alpha(c, T) + p(\psi) f_\beta(c, T) \, ,$$

$w_\phi(c) = (1 - c) \, w_A + c \, w_B$, and $f_\alpha(c, T)$ and $f_\beta(c, T)$ are taken from the regular solution model. The coefficients $\varepsilon_\psi{}^2$ and $w_\psi$ can be fixed using the phase boundary energy and interface thickness, if there exists a temperature $T_x$, where the $\alpha$ and $\beta$ phases are in equilibrium. The orientational contribution to the free energy is set as follows:

$$f_{ori}(\psi, |\nabla\theta|) = HT\{h(\psi)F_0 + [1 - h(\psi)]F_1\}$$

$$F_0 = |\nabla\theta|$$

$$F_1 = [1 - x + x|\cos(2\pi n\xi_0|\nabla\theta|)|]/(2\xi_0)$$

$$h(\psi) = \frac{1}{2}[1 + \cos(2\pi\psi)]$$

where the function $1 - h(\psi)$ activates the term $F_1$ in the interface region that initiates a jump of a prescribed amplitude in the orientation field. The following four equations of motion apply.

### A.2.1 Solid-Liquid Phase Field

$$\widetilde{\dot{\phi}} = m_\phi\left[\widetilde{\nabla}^2\phi - \xi^2\,\frac{w(c)Tg'(\phi) + p'(\phi)\{f_L(c,T) - f_S(\psi, \nabla\psi, c, T) - f_{ori}(\psi, |\nabla\theta|)\}}{\varepsilon_\phi^2 T}\right]. \quad (A7)$$

### A.2.2. Solid-Solid Phase Field

$$\widetilde{\dot{\psi}} = m_\psi[1 - p(\phi)]\left[\begin{array}{l}\widetilde{\nabla}^2\psi - \xi^2\,\dfrac{w_\psi Tg'(\psi) + p'(\psi)\{\Delta f_\alpha(c,T) - \Delta f_\beta(c,T)\}}{\varepsilon_\psi^2 T} \\[2mm] -\dfrac{H\xi^2}{\varepsilon_\psi^2}h'(\psi)[F_0 - F_1] - \dfrac{p'(\phi)}{1 - p(\phi)}\nabla\psi\nabla\phi\end{array}\right], \quad (A8)$$

where $\Delta f_{\alpha,\beta}(c,T) = f_L(c,T) - f_{\alpha,\beta}(c,T)$, and $m_\psi = M_\psi\varepsilon_\psi^2 T/D_l$ is the dimensionless mobility of the solid-solid phase field, $\psi$.

### A.2.3. Concentration Field

$$\widetilde{\dot{c}} = \widetilde{\nabla}\left\{\frac{v_m}{RT}\lambda c(1-c)\widetilde{\nabla}\left[\begin{array}{l}[1-p(\phi)]\left\{[1-p(\psi)]\dfrac{\partial f_\alpha}{\partial c}(c,T) + p(\psi)\dfrac{\partial f_\beta}{\partial c}(c,T)\right\} \\[2mm] + (w_B - w_A)Tg(\phi) + p(\phi)\dfrac{\partial f_L}{\partial c}(c,T) - \dfrac{\varepsilon_c^2 T}{\xi^2}\widetilde{\nabla}^2 c\end{array}\right]\right\}. \quad (A9)$$

### A.2.4. Orientation Field

$$\widetilde{\dot{\theta}} = m_\theta\widetilde{\nabla}\left\{[1 - p(\phi)][h(\psi) - [1 - h(\psi)]G]\frac{\widetilde{\nabla}\theta}{|\widetilde{\nabla}\theta|}\right\}, \quad (A10)$$

where $G = \text{sign}[\cos(2\pi n\xi_0|\nabla\theta|)]\sin(2\pi n\xi_0|\nabla\theta|)2\pi nx$, while the dimensionless orientational mobility is $m_\theta = [M_{\theta,S} + (M_{\theta,L} - M_{\theta,S})\cdot\phi]\,\xi HT/D_l$.

## A.3 Model C

It differs from Model A in that a new form of the orientational free energy proposed by Gránásy and Pusztai [33] is assumed, that sets preference for crystallographic branching with fixed misorientation:

$$f_{ori} = \frac{HT}{2\xi_0} \left\{ xF_0 + (1-x)F_1 \right\}$$

$$F_0 = \begin{cases} \left| \sin\left(2\pi m \xi_0 |\nabla\theta|\right) \right| & \text{for} \quad \xi_0|\nabla\theta| < \dfrac{3}{4m} \\ 1 & \text{otherwise} \end{cases}$$

$$F_1 = \begin{cases} \left| \sin\left(2\pi n \xi_0 |\nabla\theta|\right) \right| & \text{for} \quad \xi_0|\nabla\theta| < \dfrac{1}{4n} \\ 1 & \text{otherwise} \end{cases}$$

$$s(\vartheta, \theta) = 1 + s_0 \cos\left[ k(\vartheta - 2\pi\theta/k) \right], \quad \vartheta = \arctan\left[ (\nabla\phi)_y / (\nabla\phi)_x \right]$$

(A11)

The orientational free energy has two local minima as a function of the angle $\xi_0 |\nabla\theta|$, corresponding to no misorientation and a preferred misorientation (Fig. A1). This means that regions with a large enough orientation difference from a neighboring parent crystal will relax towards a finite misorientation. This selection of grain orientation only occurs provided that noise does not disrupt the process. The branching angle and the depth of this metastable minimum of $f_{ori}$ are specified by $m$, $n$ and $x$.

In any real system there will be many preferred (low energy) orientations, a reflection of the underlying crystallographic symmetries. In our illustrative calculations $n = \frac{1}{2}$ has been set, while $m = 1, 2,$ and $3$ correspond to branching with 90, 45, and 30 degrees, respectively. We note that, with appropriate choice of the parameters ($x = 0$), GFN with random orientation of the new grains [33, 99, 100, 103, 196] can also be recovered.

While the equations of motion for the phase field and concentration remain unchanged (only the actual $f_{ori}$ has to be inserted into the former), the deterministic part of the equation of motion for the orientation field takes the following form

$$\dot{\tilde{\theta}} = m_\theta \left[ \tilde{\nabla} \left\{ [1 - p(\phi)]\pi \left[ x\tilde{F}_0 m + (1-x)\tilde{F}_1 n \right] \frac{\tilde{\nabla}\theta}{|\tilde{\nabla}\theta|} \right\} - \frac{\varepsilon_\phi^2}{H\xi} s \frac{\partial s}{\partial \theta} |\tilde{\nabla}\phi|^2 \right],$$

(A12)

where

$$\tilde{F}_0 = \begin{cases} \text{sign}\left[ \sin\left(2\pi m \tilde{\xi}_0 |\tilde{\nabla}\theta|\right) \right] \cos\left(2\pi m \tilde{\xi}_0 |\tilde{\nabla}\theta|\right) & \text{for} \quad \tilde{\xi}_0|\tilde{\nabla}\theta| < \dfrac{3}{4m} \\ 0 & \text{otherwise} \end{cases}$$

$$\tilde{F}_1 = \begin{cases} \text{sign}\left[ \sin\left(2\pi n \tilde{\xi}_0 |\tilde{\nabla}\theta|\right) \right] \cos\left(2\pi n \tilde{\xi}_0 |\tilde{\nabla}\theta|\right) & \text{for} \quad \tilde{\xi}_0|\tilde{\nabla}\theta| < \dfrac{1}{4n} \\ 0 & \text{otherwise} \end{cases},$$

while $\tilde{\xi}_0 = \xi_0 / \xi$ is the dimensionless correlation length of the orientation field.

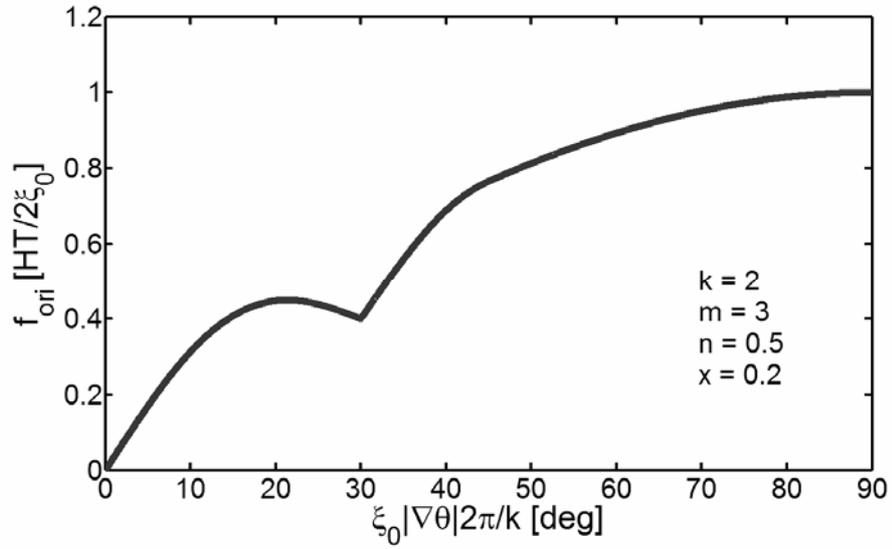

**Figure A1.** Orientational free energy $f_{ori}$ in Model C, as a function of misorientation angle (in degree) for two-fold symmetry ($k = 2$), while $n = \frac{1}{2}$, $m = 3$, and $x = 0.2$. If the neighboring pixel has a smaller misorientation than ~20° (local maximum), it can reduce the free energy by relaxing to the bulk crystal orientation (0°). If misorientation is larger than this, the closest minimum is 30°. So, neighboring pixels of large misorientation tend to relax to 30°, unless fluctuations prevent this. Note that $\theta$ is an angular variable, so the maximum possible misorientation is $\Delta\theta_{max} = 0.5$.